\documentclass[runningheads]{llncs}
\usepackage[T1]{fontenc}
\usepackage{graphicx}
\usepackage{listings}
\usepackage{booktabs}
\usepackage{mathtools}
\usepackage{tikz}
\usepackage{pgfplots}
\usepackage[colorlinks,naturalnames,citecolor=blue,urlcolor=blue]{hyperref}
\newcommand{\myhref}[1]{\href{#1}{\url{#1}}}
\usepackage{xspace}
\usepackage{adjustbox}
\usepgfplotslibrary{groupplots}
\usepackage[normalem]{ulem}
\usepackage{subcaption}

\usepackage{color}

\newcommand{\TLA}{TLA\textsuperscript{+}\xspace}

\newcommand{\cvc}{\texttt{cvc5}\xspace}
\newcommand{\nuxmv}{\texttt{nuXmv}\xspace}
\newcommand{\ltlsynt}{\texttt{ltlsynt}\xspace}
\newcommand{\qwen}{\textsc{Qwen}\xspace}
\newcommand{\gpt}{\textsc{GPT-5}\xspace}

\newcommand{\qwenAIGER}{$\textsc{Qwen}^{\text{AIG}}$\xspace}
\newcommand{\gptAIGER}{$\textsc{GPT-5}^{\text{AIG}}$\xspace}
\newcommand{\qwenSMV}{$\textsc{Qwen}^{\text{SMV}}$\xspace}
\newcommand{\gptSMV}{$\textsc{GPT-5}^{\text{SMV}}$\xspace}

\newcommand{\qwenrec}{\textsc{Qwen}\xspace}

\newcommand{\qwendists}{\textsc{Qwen}\xspace}

\newcommand{\gptrec}{\textsc{GPT-5}\xspace}
\newcommand{\gptdists}{\textsc{GPT-5}\xspace}
\newcommand{\gptdistww}{\textsc{GPT-5-NoGram}\xspace}

\newcommand{\cataclyst}{\textsc{Cataclyst}\xspace}
\newcommand{\psem}{\textsc{PolySemist}\xspace}

\let\oldsubsubsection\subsubsection
\renewcommand{\subsubsection}[1]{\oldsubsubsection{#1.}}

\definecolor{systemBg}{RGB}{240, 242, 245}
\definecolor{systemBorder}{RGB}{200, 210, 220}
\definecolor{userBg}{RGB}{235, 245, 250}
\definecolor{userBorder}{RGB}{180, 210, 230}
\definecolor{promptComment}{RGB}{100, 100, 100}
\definecolor{promptKeyword}{RGB}{0, 100, 150}
\definecolor{promptString}{RGB}{160, 70, 160}

\lstdefinestyle{systempromptSimple}{
    basicstyle=\small\ttfamily,
    breaklines=true,
    breakatwhitespace=false,
    keepspaces=true,
    showspaces=false,
    showstringspaces=false,
    showtabs=false,
    tabsize=2,
    captionpos=t,
    title={\textbf{System Prompt}},
}

\lstdefinestyle{userpromptSimple}{
    basicstyle=\small\ttfamily,
    breaklines=true,
    breakatwhitespace=false,
    keepspaces=true,
    showspaces=false,
    showstringspaces=false,
    showtabs=false,
    tabsize=2,
    captionpos=t,
    title={\textbf{User Prompt}},
}

\lstdefinestyle{systemprompt}{
    backgroundcolor=\color{systemBg},
    frame=single,
    framesep=8pt,
    framexleftmargin=12pt,
    framexrightmargin=6pt,
    framextopmargin=6pt,
    framexbottommargin=6pt,
    rulecolor=\color{systemBorder},
    basicstyle=\small\ttfamily,
    commentstyle=\color{promptComment}\itshape,
    keywordstyle=\color{promptKeyword}\bfseries,
    stringstyle=\color{promptString},
    breaklines=true,
    breakatwhitespace=false,
    keepspaces=true,
    showspaces=false,
    showstringspaces=false,
    showtabs=false,
    tabsize=2,
    captionpos=t,
    title={\textbf{System Prompt}},
    morekeywords={INSTRUCTION, CONTEXT, EXAMPLES, TASK, NOTE},
}

\lstdefinestyle{userprompt}{
    backgroundcolor=\color{userBg},
    frame=single,
    framesep=8pt,
    framexleftmargin=12pt,
    framexrightmargin=6pt,
    framextopmargin=6pt,
    framexbottommargin=6pt,
    rulecolor=\color{userBorder},
    basicstyle=\small\ttfamily,
    commentstyle=\color{promptComment}\itshape,
    keywordstyle=\color{promptKeyword}\bfseries,
    stringstyle=\color{promptString},
    breaklines=true,
    breakatwhitespace=false,
    keepspaces=true,
    showspaces=false,
    showstringspaces=false,
    showtabs=false,
    tabsize=2,
    captionpos=t,
    title={\textbf{User Prompt}},
    morekeywords={QUERY, INPUT, QUESTION, PROBLEM},
}

\begin{document}
\title{Can LLMs Perform Synthesis?}
%


\author{Derek Egolf
\and
Yuhao Zhou
\and
Stavros Tripakis
}
\authorrunning{D. Egolf et al.}
%
\institute{Northeastern University, Boston MA, USA\\
\email{\{egolf.d,zhou.yuhao,stavros\}@northeastern.edu}
}
\maketitle              
%
\begin{abstract}
How do LLMs compare with symbolic tools on program synthesis tasks? We
investigate this question on several synthesis domains: LTL reactive synthesis,
syntax-guided synthesis, distributed protocol synthesis, and recursive function
synthesis. For each domain, we choose a state-of-the-art symbolic tool and
compare it to 
an open-source, 32 billion parameter version of the \qwen LLM
and the proprietary, frontier LLM \gpt. We
couple \qwen with a symbolic verifier and run it repeatedly until it either
produces a solution that passes the verifier, or until there is a timeout, for
each benchmark. We run \gpt once per benchmark and verify the generated output.
In all domains, the symbolic tools solve more benchmarks than \qwen and either
outperform or are about on par with \gpt. In terms of execution time, the
symbolic tools outperform \gpt in all domains, and either outperform or are
very close to \qwen, despite the fact that the LLMs are run on significantly
more powerful hardware.

\end{abstract}
%
%
%


\section{Introduction}

The goal of program synthesis has been to generate, as automatically as possible, systems that are {\em correct by construction}, that is, that are guaranteed to satisfy a given correctness specification~\cite{Church57,Finkbeiner2016NATO}.
Since the publication of Church's problem in the 1950s~\cite{Church57}, synthesis has come a long way, from the early approaches of deductive program synthesis~\cite{MannaWaldinger1980} and reactive synthesis~\cite{PnueliRosner89,Finkbeiner2016NATO}, to more recent approaches such as sketching~\cite{Solar-LezamaPLDI2005,ArmandoSTTT2013}, syntax-guided synthesis~\cite{Sygus13}, and others~\cite{GulwaniPolozovSingh2017}.

On the other hand, the field of programming itself is currently undergoing rapid changes;
LLMs are now routinely used to generate code for 
common programming tasks~\cite{openai2025competitiveprogramminglargereasoning,chen2021humanEval,zhuo2024bigcodebench}.

These developments lead naturally to the question: {\em can LLMs do synthesis?}
Specifically, we want to evaluate how good LLMs are at synthesizing
programs from formal specifications, compared to symbolic tools.
Therefore, we rephrase our research question as follows: {\em how do LLMs compare with state-of-the-art symbolic synthesis tools?}

\subsubsection{Choice of synthesis domains}

As there are many kinds of synthesis problems, the answer to the above question generally depends on the particular synthesis domain that is chosen.
In this paper, we examine the following synthesis domains:
\begin{itemize}
\item LTL reactive synthesis~\cite{PnueliRosner89,Finkbeiner2016NATO}: the problem is to synthesize a finite state machine (FSM) which implements ({\em realizes}) a given specification in Linear Temporal Logic (LTL)~\cite{PnueliLTL77}.

\item Syntax-guided synthesis (SyGuS)~\cite{Sygus13}: the problem is to synthesize, given a specification and a grammar, a program that can be generated by the grammar and that satisfies the specification.

\item Distributed protocol synthesis by sketching~\cite{egolfFMCAD2024,egolf2025tacas}: given a sketch of a distributed protocol in \TLA\cite{tla-lang}, the problem is to complete the sketch such that the resulting protocol satisfies a given set of \TLA properties.
\item Recursive program synthesis by sketching~\cite{egolf2026recursiveprogramsynthesissketches}: given a sketch of a recursive function in the LISP-like ACL2s programming language~\cite{acl2s}, and a specification in first-order logic, the problem is to complete the sketch such that the resulting program satisfies the specification.
\end{itemize}

\subsubsection{Choice and use of LLMs}

Across all domains, we compare the corresponding symbolic tool with two LLMs:
  \textsc{Qwen-2.5-Coder-32B-Instruct}~\cite{hui2024qwen25coder}
(hereafter referred to as \qwen) and 
\gpt~\cite{singh2025gpt5systemcard}. \qwen is a 32 billion parameter
model. To our knowledge, it is the state-of-the-art open-source LLM for
code generation that fits on an NVIDIA H200 GPU (the hardware that we use to run it). \gpt is a proprietary
model, and to our knowledge OpenAI has not publicly disclosed the number of
parameters it uses. GPT-3, a model released in 2020, had 175 billion
parameters~\cite{brown2020gpt3}, so we can reasonably assume that \gpt has
significantly more than that.

We emphasize that, in all cases, we use these LLMs ``out of the box'' meaning that we do not train the LLMs ourselves, we do not fine-tune them, we do not use RAG or similar techniques, and so on. 
In the spirit of our original question {\em can LLMs do synthesis?}, we want to evaluate to what extent LLMs are able to do synthesis {\em on their own}, as opposed to how LLMs might be integrated into more sophisticated ``hybrid'', ``neurosymbolic'', or similar techniques. We do believe that such techniques are important, but they are beyond the scope of our comparison here.
(We review related neurosymbolic techniques in our related work section \S\ref{sec:related}.)

At the same time, we take a pragmatic, user-oriented perspective. How might a
user use an LLM for code synthesis? LLM outputs come with no guarantees, so a
user will have to check the output to ensure that it satisfies the problem
requirements. For that reason, we couple LLMs with a symbolic {\em verifier}
which checks the LLM output, and potentially has the LLM retry in case of
failure. In the spirit of fairness, we try to allocate similar resources to the
toolchains under comparison, as much as possible. 

Section~\ref{sec_common} that follows presents in more detail the way we use LLMs, discusses fairness considerations and resource allocation, and describes other common elements across all domains.
Sections~\ref{sec_reactive_synth}-\ref{sec_acl2s_synth}  present the experimental results for each domain.
Section~\ref{sec:related} discusses related work.
Section~\ref{sec_conclusion} concludes the paper.

\section{Methodology and Common Elements Across Domains}
\label{sec_common}

\ifdefined\onlyinfullpaper
\begin{problem}[Code Generation]
    Given a natural language description of a programming task, generate code
    that satisfies the description. The code is evaluated based on (typically
    hidden) input-output examples.
\end{problem}

\begin{problem}[Formal Synthesis]
    Given a formal specification $\phi$ describing a programming task and a
    formal satisfaction relation $\models$, generate code $c$ such that $c
    \models \phi$. The implementation is evaluated based on whether it
    satisfies the (observable) specification.
\end{problem}
\fi

\subsubsection{Fair comparison}

A fair comparison of two tools ideally requires running both tools on the same
computing hardware and allocating the same time and memory resources to both.
This is impossible to achieve in our case, because, unlike all the symbolic
synthesis tools that we consider, neither \qwen nor \gpt can be effectively run on a
CPU. As we still want our comparison to be as fair as possible, our
philosophy is to provide as equal as possible computational resources to the
compared toolchains. By {\em computational resources} we mean the product of
{\em hardware computing power} and {\em time allocated to perform the
computation} (which we control by means of {\em timeouts}). Ideally, we would
like this product to be equal so that, for example, if tool A is run on a
hardware twice as fast as that of tool B, then the timeout for tool B should be
two times that of tool A. In practice, however, this calculation is impossible
to achieve, because the computing power of hardware is difficult to
measure/compare across hardware architectures.
Moreover, to our knowledge, OpenAI does not even disclose the exact
hardware used to run \gpt.

We therefore take a pragmatic approach, and proceed as follows:
\begin{itemize} 
\item We run all symbolic tools on CPUs, with a 10-minute timeout for each benchmark
(15 mins for the recursive program synthesis domain in \S\ref{sec_acl2s_synth}).
\item We use a single NVIDIA H200 GPU to run \qwen.
We couple \qwen with a verifier (run on a CPU)
which checks each candidate that \qwen generates. For each benchmark, the whole
process is iterated until a correct solution is found, or a 10-minute time
limit is reached (15 min for \S\ref{sec_acl2s_synth}). 
\item We run \gpt using the OpenAI API. In all domains, we set an output token budget of
16,384 and set the reasoning parameter to ``medium''. 
 For each benchmark, we run \gpt once, and check any generated
solution with a verifier (run on a CPU), allocating a total of 10
minutes both for \gpt generation and verification (15 min for \S\ref{sec_acl2s_synth}).
\end{itemize}
The rationale is to give the same timeout to each toolchain, even though the
toolchains are executed on different hardware. To compensate for the fact that
\gpt is itself very powerful and also run on very powerful hardware, we only
query \gpt once per benchmark, and do not iterate like we do with \qwen. This
is still a somewhat unfair comparison between symbolic tools and \qwen, as the
latter is run on much more powerful hardware. 
We accept this until a better comparison methodology is found. 

\subsubsection{LLM toolchains}

\begin{figure}
    \centering
    \resizebox{\textwidth}{!}{%
    \begin{tikzpicture}[
        box/.style={rectangle, draw, minimum width=2cm, minimum height=1cm, align=center},
        arrow/.style={->, >=stealth, thick}
    ]
        \node[box] (box1) at (0,0) {Prompt\\ Constructor};
        \node[box] (box2) at (4,0) {LLM};
        \node[box] (box3) at (8,0) {Post\\ Processor};
        \node[box] (box4) at (12,0) {Verifier};
        
        \draw[arrow] (-2.5,0) -- (box1) node[above, pos=0.5, align=center] {formal\\ spec.} node[below, pos=0.5] {$\Phi$};
        \draw[arrow] (box1) -- (box2) node[above, pos=0.5] {prompt};
        \draw[arrow] (box2) -- (box3) node[above, pos=0.5, align=center] {raw\\ response};
        \draw[arrow] (box3) -- (box4) node[above, pos=0.5, align=center] {formal\\ model} node[below, pos=0.5] {$C$};
        \draw[arrow] (box4) -- (14.5,0) node[above, pos=0.5, align=center] {return C} node[below, pos=0.5] {$C\models\Phi$};
        \draw[arrow] (box4.south) -- (12,-0.75) node[right, pos=1] {$C\not\models\Phi$} -- (2.8,-.75) node[below, pos=0.5] {reprompt} -- (2.8,0);
    \end{tikzpicture}%
    }
    \caption{LLM-based synthesis methodology: the \qwen toolchain has the {\em reprompt} loop, but the \gpt toolchain does not (\gpt is only queried once per benchmark).}
    \label{fig:llm-method}
\end{figure}
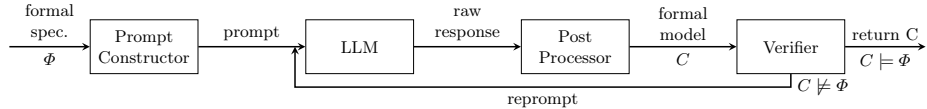

Pictorially, our LLM toolchains across all domains look like the one in Figure~\ref{fig:llm-method}.
Note that the \qwen toolchain has the {\em reprompt} loop, but the \gpt toolchain does not (\gpt is only queried once per benchmark).
We handle timeouts as follows.
In the \gpt toolchain, we allocate a 10-min time limit (15 min for \S\ref{sec_acl2s_synth}) to the entire toolchain, per benchmark. This means, for example, that if \gpt takes 3 mins to generate a candidate solution, then the verifier has 7 mins to check that candidate. 
A solution which is both generated and verified correct within 10 mins is counted as a correct solution.
In the \qwen toolchain, we allocate 10 mins per benchmark (15 min for
\S\ref{sec_acl2s_synth}), and let the loop iterate as many times as possible until the 10 mins are exhausted, or a correct solution is found. Again, we count a solution as correct if it is both generated and verified within the time limit.

We call this method {\em iterate-LLM-until-solution-or-timeout} (ILST). We
remark that, for a given benchmark, the prompt remains the same for each
iteration of the loop. That is, we do not provide any verification feedback to
the LLM: no counterexamples and no other indication that the previously
generated solutions were incorrect. This would not only require adding the
previously generated (incorrect) solutions to the prompt, but it would also go
against our premise to test whether LLMs can do synthesis {\em on their own},
and not as part of a neurosymbolic loop. It therefore often happens that the
LLM generates the same solution many times during the loop, despite using
temperature and other parameters which make the generation more flexible
and less deterministic (for all \qwen toolchains we use temperature=0.8,
top\_p=0.95, and top\_k=50).
In some domains, we cache the generated solutions within the ILST loop, and if
the same solution is generated twice, we skip the verification step since we
already know that this solution is incorrect.

We remark that we also considered the option of asking the LLM to generate a
bundle of, say $k>1$, different solutions, and then verifying all of them. We
opt instead to use ILST for the following reasons. First, 
in the $k$-bundle approach, it is unclear how to handle verification time budgets in a meaningful way. For example, what happens if the LLM generates $k=2$ solutions and the first one is model-checked to be
incorrect while the second one is model-checked correct but the model-checker
exceeds its timeout?
In contrast, in ILST, the time spent generating and verifying is automatically balanced appropriately.
Second, 
ILST more closely resembles
how a human would use the LLM: generate a solution, check it, and try again if
necessary. Finally, many symbolic synthesis tools are {\em enumerative} in
nature, meaning that they enumerate a candidate solution, check it, and if
incorrect, try again. This similarity makes ILST a natural fit for comparison
against symbolic tools.

\section{LTL Reactive Synthesis}
\label{sec_reactive_synth}

The problem is to synthesize an FSM which implements ({\em realizes}) a given LTL specification~\cite{PnueliRosner89,Finkbeiner2016NATO}.
We adopt the problem format from the LTL Synthesis Track of SYNTCOMP~\cite{jacobs2024SYNTCOMP} -- \myhref{https://www.syntcomp.org/}.
SYNTCOMP uses the Temporal Logic Synthesis Format (TLSF)~\cite{jacobs2016tlsfLanguage} for specifications. 
A TLSF file declares the input and output variables of the FSM to be synthesized, and the LTL formulas over these variables that the FSM must implement.
\ifdefined\onlyinfullpaper
Figure~\ref{fig:tlsf_example} illustrates a simple example: an AND gate whose output must always equal the conjunction of its two inputs.

\begin{figure}[t]
\centering
\begin{lstlisting}[basicstyle=\ttfamily\small]
INFO {
    TITLE:       "Trivial AND gate"
    DESCRIPTION: "Output is AND of two inputs"
    SEMANTICS:   Mealy
    TARGET:      Mealy
}
MAIN {
    INPUTS { a; b; }
    OUTPUTS { outp; }
    GUARANTEES { G(outp <-> (a && b)); }
}
\end{lstlisting}
\caption{A TLSF specification for an AND gate.}
\label{fig:tlsf_example}
\end{figure}

The synthesis goal is to construct a finite-state machine that realizes the specification, accepting inputs and producing outputs such that all traces satisfy the LTL guarantees.

\begin{problem}
\label{prob:ltlSynth}
Given an LTL formula $\varphi$ over finite sets of input and output variables, find a finite-state machine $M$ over the same input and output alphabets such that $M \models \varphi$.
\end{problem}
\fi

\subsection{Experimental Setup}

\subsubsection{Symbolic tool}
We compare LLMs with the state-of-the-art synthesis tool \ltlsynt~\cite{renkin2022ltlsynt}, which solved the most benchmarks in the LTL Synthesis Track of SYNTCOMP 2025.
\ltlsynt produces solutions in the AIGER format~\cite{biere2007aiger}.
We allocate a 10-minute timeout to \ltlsynt for synthesis.
We adhere to SYNTCOMP's verification methodology and employ its verification pipeline to check all results of \ltlsynt for correctness. 
First, we use \texttt{SyFCo}~\cite{jacobs2016tlsfLanguage} to convert the TLSF input into SMV/\nuxmv format, which is subsequently compiled into an AIGER monitor.
This monitor is then combined with the candidate solution, and the combined AIGER file is checked by the \nuxmv model checker~\cite{cavada2014nuxmv}.
We allocate a 10-minute timeout for this entire verification pipeline.
Such post-synthesis verification is unnecessary in principle for a symbolic tool like \ltlsynt which guarantees correctness by construction. However, we still  
perform it in order to ensure correctness and maintain consistency with the LLM experiments.

\subsubsection{LLM Toolchains}

As the choice of output format can impact LLM performance, for this domain, we ask each LLM to synthesize the FSM either in AIGER~\cite{biere2007aiger} (the output format of SYNTCOMP) or in SMV (the input format of \nuxmv). \qwenAIGER and \gptAIGER (respectively, \qwenSMV and \gptSMV) denote the LLM toolchains that generate AIGER (respectively, SMV) outputs. The AIGER toolchains use different prompts and verifiers than the SMV ones.

For \qwen toolchains we use an ILST loop (c.f. \S\ref{sec_common}) with a cumulative 10-minute timeout, while for \gpt toolchains we use a single-pass generation with a combined 10-minute timeout for generation and verification. For \qwen toolchains we do not impose a token limit, allowing the model to generate as much output as needed within the time constraint. For \gpt toolchains we enforce a combined 10-minute timeout for generation and verification.

For verification, the AIGER toolchains use the same verification procedure as \ltlsynt's post-synthesis verification described above. 
Any syntactic errors, such as malformed AIGER files or inconsistencies between the solution's input/output symbols and the TLSF specification, are automatically detected during this verification procedure. 
In the SMV toolchains, we convert the TLSF specifications to SMV format using \texttt{SyFCo}, combine them with the LLM-generated SMV module, and model-check the result with \nuxmv \texttt{-ctt}. The \texttt{-ctt} flag ensures that the transition relation is total (this, for instance, prevents the LLM from ``cheating'' by generating solutions that trivially satisfy the LTL formulas by deadlocking). We remark that FSMs written in AIGER are deterministic by construction, but LLMs that output SMV could theoretically generate nondeterministic FSMs (but after carefully crafting the prompt, this ends up never happening in our experiments). 
We therefore check that the results of the SMV LLM toolchains are deterministic. We perform this check only post-synthesis, and not during the ILST loop, because checking determinism is computationally expensive (see Appendix~\ref{app:ltl_synth_sanity}). Inside the loop, we rely solely on \nuxmv \texttt{-ctt}.

\subsubsection{Prompts}

For the AIGER LLM toolchains, the prompt instructs the LLM to synthesize an implementation of the input TLSF specification in the ASCII AIGER format. The prompt consists of a brief description of the ASCII AIGER format followed by illustrative examples that demonstrate expected LLM's output in AIGER. 
The complete prompt is provided in Appendix~\ref{app:ltl_synth_prompt_AIGER}.

The prompt for the SMV toolchains is designed to ensure that the LLM generates a valid, deterministic FSM. To achieve this, we impose strict structural constraints. First, a rigid mapping between TLSF concepts and SMV sections: (1) all TLSF inputs must be declared in an \texttt{IVAR} section to prevent assignments to input variables (which can only be controlled by the environment); (2) internal state variables must be declared in the \texttt{VAR} section; (3) all variables must be of type \texttt{boolean}; (4) TLSF outputs must be defined in the \texttt{DEFINE} section. 
Second, to avoid nondeterministic solutions, we request that the FSM is defined using only an \texttt{ASSIGN} block with explicit \texttt{init()} and \texttt{next()} assignments for every state variable, and we forbid the use of \texttt{INIT} and \texttt{TRANS} sections. We also provide explicit instructions on boolean operator syntax to minimize syntax errors during generation. 
Finally, as none of the benchmarks in our suite require a Moore target FSM (where outputs depend exclusively on state), we do not add any relevant instructions in our prompt.
The complete prompt is provided in Appendix~\ref{app:ltl_synth_prompt_SMV}.

\subsubsection{Benchmarks}
We use benchmarks from the LTL Synthesis Track of SYNTCOMP 2025.
There are 1586 benchmarks in that suite.
Some of these benchmarks are labeled as realizable, unrealizable, or unknown.
For our experiments, we select all benchmarks marked as realizable, a total of {433} benchmarks.
We only select realizable benchmarks in order to avoid confusing the LLMs.

\subsubsection{Hardware and Parameters}
We ran \ltlsynt on a machine with 4 CPU cores and 64GB RAM. The \qwen toolchains ran with the same hardware specification plus an NVIDIA H200 GPU. We ran \gpt via official OpenAI API.

\subsection{Experimental Results}
\label{sec_reactive_synth_results}

The results are presented in Table~\ref{tab:ltl_results}.
The {\em Solved} column reports the number of correct solutions obtained within the 10-minute timeout.
\ltlsynt generated solutions for 396 out of the 434 benchmarks. 
We were able to verify (post-synthesis) 354 of those solutions; for the remaining 42, \nuxmv timed out after 10 minutes (this 10-minute limit applies only to post-synthesis verification and is not counted towards the 10-minute timeout allocated to \ltlsynt).
In the $433-396=37$ benchmarks that \ltlsynt did not solve, it timed out 34 times (generating no solution) and crashed 3 times (e.g., due to memory allocation failure).

\ltlsynt solved 143 benchmarks that no LLM toolchain managed to solve.
\gptAIGER solved 2 benchmarks that \ltlsynt and the other LLM toolchains failed to solve.
Similarly, \gptSMV solved 3 benchmarks that no other method could solve.
We also observed that \ltlsynt solved every benchmark that the \qwen toolchains successfully solved, while \qwenAIGER and \qwenSMV solved 1 and 6 benchmarks that \gptAIGER and \gptSMV failed to solve, respectively.
In total, 401 out of 434 benchmarks were solved by at least one method.

\begin{table*}
\centering

\begin{tabular}{l|c|ccc|cccc|cccc}
\hline
\textbf{Method} & \textbf{\# Solved} & \multicolumn{3}{|c|}{\textbf{Success Time}} & \multicolumn{4}{|c|}{\textbf{Success Iterations}} & \multicolumn{4}{c}{\textbf{Fail Iterations}} \\
\cline{3-5} \cline{6-8} \cline{9-13}
 & within TO & Min & Max & Mean & Min & Max & Mean & Med & Min & Max & Mean & Med \\
\hline
\ltlsynt    & 354+42 & 0.5 & 301.6 & 4.6   & --   & --   & --  & --  & --  & --    & -- & --   \\
\qwenAIGER  &   4       & 5.4  & 261.5 &  79.6 & 5    & 136  & 39.2 & 8  & 2   & 522   & 41.5 &  7 \\
\qwenSMV    &  50       & 1.5  & 575.9 & 164.9 & 1    & 135  & 23.9 & 9  & 1   & 202   & 42.1 & 33 \\
\gptAIGER   & 150       & 14.9 & 311.4 & 124.0 & --   & --   & --   & --   & --  & --    & --   & --   \\
\gptSMV     & 229       & 13.8 & 465.3 & 107.0 & --   & --   & --   & --   & --  & --    & --   & --   \\
\hline
\end{tabular}%
\caption{Summary statistics on 433 LTL reactive synthesis benchmarks.\label{tab:ltl_results}}
\end{table*}

The {\em Success Time} columns present execution time statistics (minimum, maximum, and average execution time in seconds) only for the successful cases, where the toolchains generate a correct solution.
For \ltlsynt, we count all 396 generated outputs as correct solutions for the purpose of execution time statistics.
As shown, \ltlsynt significantly outperforms the LLM toolchains in the execution time metric.

The {\em Success Iteration} columns present the minimum, maximum, average, and median number of iterations required for the \qwen ILST loops to find a solution (successful cases only). \qwenSMV sometimes generates a correct solution on the first try, while in other cases it requires up to 135 attempts (recall that no feedback is provided to the LLM).
The {\em Fail Iteration} columns present the same statistics for the failed cases. The minimum of 1 for \qwenSMV occurs because, in some cases, \nuxmv timed out while verifying the very first candidate produced. 

\subsubsection{Determinism Checking of LLM SMV Solutions}
As discussed above, we verify all LLM SMV solutions post-synthesis, to ensure that they represent deterministic FSMs.
We checked the determinism of all solved LLM-generated SMV benchmarks of Table~\ref{tab:ltl_results} automatically, subject to a 10-minute timeout per benchmark.
We used a standard {\em self-composition} technique which consists in instantiating two copies of the FSM, feeding the same inputs to both, and checking that in all reachable global states, the two FSMs have identical local states
(see Appendix~\ref{app:ltl_synth_sanity}).
All 50 \qwenSMV solutions were verified as deterministic within the timeout.
Out of the 229 \gptSMV solutions, 225 were verified automatically as deterministic.
In the remaining 4 cases the automatic determinism check timed out, but we confirmed the determinism of all these 4 solutions through manual inspection.

\subsubsection{LLM Response Failures}

\qwenSMV solved 50 benchmarks and reached the 10-min timeout in the remaining $433-50=383$ cases.
In 276 of these 383 cases, a final candidate was generated by the LLM, but was not model-checked (due to the timeout having been reached).
Since 276 is a large percentage of 383, we used \nuxmv to check whether any of those ``last-minute'' solutions happened to be correct.
All 276 turned out to be wrong: 134 were syntactically incorrect (i.e., files that \nuxmv could not parse)
and 142 were semantically incorrect (i.e., violating the LTL formulas).

\gptAIGER ran out of its token budget in 90 cases, generating no solution.
In all remaining $433 - 90 = 343$ cases, it generated a solution within the 10-min timeout.
These 343 candidates were then verified by \nuxmv (with the timeout set to the remaining time within the total 10-minute limit).
Of these 343 candidates, 150 passed the model checker (correct solutions);
145 failed due to syntax errors;
48 failed due to semantic errors;
and no cases timed out during model checking.

\gptSMV exhausted its token budget in 72 cases.
In all remaining $433 - 72 = 361$ cases, it generated a solution within the 10-minute time limit.
These 361 candidates were then verified by \nuxmv (with the timeout set to the remaining time within the total 10-minute limit).
Of these 361 candidates, 229 passed the model checker (correct solutions);
51 failed due to syntax errors;
40 failed due to semantic errors;
and 41 timed out during model checking. 
We reran \nuxmv on the latter 41 cases with a fresh 10-minute timeout, but all still timed out during model checking
(recall that for \gpt we use a total 10-minute timeout for both generation and verification, so the original \nuxmv runs generally had less than 10 minute available).

\subsubsection{\gpt: Tokens and Monetary Cost}

For the \gptAIGER experiment, the average input length was 1,831 tokens per benchmark, and the average output length was 10,004 tokens. The total cost for all 433 queries was 43.97 USD.
For the \gptSMV experiment, the average input length was 1,737 tokens per benchmark, and the average output length was 8,930 tokens. The total cost for all 433 queries was 39.58 USD.

\section{Syntax-Guided Synthesis}
\label{sec_sygus_synth}

The SyGuS problem is to synthesize, given a specification and a grammar, a program that (1) can be generated by the grammar and (2) satisfies the specification~\cite{Sygus13}.
We adopt the SyGuS Input Format (SyGuS-IF) 2.1 standard~\cite{padhi2023sygusif2} for expressing SyGuS problems. A SyGuS-IF problem description consists of three main components: a background theory (e.g., linear integer arithmetic), one or more function signatures to synthesize along with optional grammar constraints, and a set of semantic constraints that the synthesized functions must satisfy.

\ifdefined\onlyinfullpaper
\begin{figure}[t]
\centering
\begin{lstlisting}[language=lisp,basicstyle=\ttfamily\small]
(set-logic LIA)
(synth-fun max ((x Int) (y Int)) Int
  ((Start Int) (StartBool Bool))
  ((Start Int (0 1 x y
               (+ Start Start)
               (- Start Start)
               (ite StartBool Start Start)))
   (StartBool Bool ((and StartBool StartBool)
                    (not StartBool)
                    (<= Start Start)))))
(synth-fun min ((x Int) (y Int)) Int)
(declare-var x Int)
(declare-var y Int)
(constraint (>= (max x y) x))
(constraint (>= (max x y) y))
(constraint (or (= x (max x y)) (= y (max x y))))
(constraint (= (+ (max x y) (min x y)) (+ x y)))
(check-synth)
\end{lstlisting}
\caption{An example SyGuS specification for synthesizing \texttt{max} and \texttt{min} functions over integers.}
\label{fig:sygus_example}
\end{figure}

Figure~\ref{fig:sygus_example} illustrates a representative SyGuS specification. The \texttt{synth-fun} declarations define the functions to be synthesized: \texttt{max} includes an explicit grammar restricting the solution to expressions built from constants, variables, arithmetic operations, and conditionals, while \texttt{min} permits any expression in the background theory. The grammar is specified using non-terminals (\texttt{Start} for integer expressions and \texttt{StartBool} for Boolean expressions) and their associated production rules. The semantic constraints collectively specify that \texttt{max} returns the larger of its inputs and that the sum of \texttt{max} and \texttt{min} equals the sum of the inputs.

The goal of synthesis is to find implementations for the declared functions that satisfy both the semantic constraints and the syntactic restrictions imposed by the grammars. 

\begin{problem}
\label{prob:sygusSynth}
Given the correctness specification $\varphi$ and the grammar $G$ defining the syntactic constraints, find an implementation $e$ such that $e \models \varphi$ and $e$ conforms to $G$.
\end{problem}
\fi

\subsection{Experimental Setup}

\subsubsection{Symbolic tool}

We compare LLMs with the state-of-the-art tool \cvc~\cite{barbosa2022cvc5}.
Its precursor \texttt{cvc4}~\cite{barrett2011cvc4} was the winner of 4/5 tracks of the last SyGuS
Competition~\cite{alur2019sygusComp2018}, held in 2019.
Given a SyGuS-IF 2.1 specification, \cvc. attempts to synthesize function implementations in SMT-LIB format~\cite{barrett2010smtLib}.
We use \cvc in its default setting, without supplying any additional command line arguments to it.
For this synthesis task, we allocate a 10-minute timeout to \cvc.

We verify the solutions produced by \cvc for both grammatical and semantic correctness.
(In principle this is unnecessary, as \cvc guarantees correctness by construction. However, we still perform this post-synthesis verification to ensure correctness and maintain consistency with the LLM experiments.)
First, a custom grammar checker ensures that the generated solutions conform to any syntactic constraints defined in the original SyGuS-IF specification (if no grammar is specified, this check trivially passes).
Second, to verify semantic correctness, we use a custom tool to convert the SyGuS-IF specification into an SMT query; we then combine this query with the synthesized function implementations and verify it using the Z3 solver~\cite{de2008z3}.
We rely on custom tools for these verification steps because the standard utilities from SyGuS-COMP 2019 can only process SyGuS-IF 1.0 files, whereas \cvc and our benchmarks adhere to SyGuS-IF 2.1.
The timeout for these post-synthesis verification tasks is set to 10 minutes per benchmark.

\subsubsection{LLM toolchains}

We evaluate a \qwen ILST loop and a \gpt toolchain, as described in \S\ref{sec_common}.
Both are instructed to generate SMT-LIB output.
In both cases we apply the same verification procedure as for \cvc: solutions are checked for both grammar conformance and semantic correctness.
A solution is {\em correct} if and only if it passes both checks.
In the \qwen toolchain, the ILST loop iterates until a correct solution is found, 
subject to a cumulative timeout of 10 minutes per benchmark (with no token limit).
In the \gpt case we generate a single candidate solution per benchmark. This candidate is then verified for both semantic correctness and grammar conformance.
We use a total 10-min timeout for the entire generation and verification process.

\subsubsection{Prompts}

Both the \qwen and \gpt toolchains are provided with an identical prompt.
The prompt instructs the model to generate SMT-LIB function implementations that satisfy the grammar and semantic constraints of the SyGuS-IF input.
We strictly constrain the output format: the LLM must produce only self-contained \texttt{(define-fun ...)} expressions that match the signature of the target \texttt{synth-fun}, without helper functions or natural language explanations.
To prevent common syntax errors, the prompt explicitly lists valid SMT-LIB primitives.
For instance, it enforces the use of \texttt{ite} for conditionals (rejecting \texttt{if}), and requires correctly typed bitvector literals (e.g., \texttt{(\_ bv10 32)}). 
The full prompt is provided in Appendix~\ref{app:sygus_synth_prompt}.

\subsubsection{Benchmarks}

We select our benchmarks from the \texttt{test/regress/cli} directory in the \cvc repository~\cite{cvc5repository}, which contains more recent problems than the SyGuS Competition (SyGuS-COMP)~\cite{alur2019sygusComp2018}, last held in 2019. We extract from the above directory all realizable SyGuS-IF specifications, excluding those requiring optional SyGuS-IF features (c.f. Appendix~C in~\cite{padhi2023sygusif2}) or experimental \texttt{cvc5} features. This extraction yields 148 benchmarks in total.

\subsubsection{Hardware and Parameters}
We ran \cvc on a machine with 4 CPU cores and 64GB RAM. The \qwen toolchain ran with the same hardware specification plus an NVIDIA H200 GPU. We ran \gpt via official OpenAI API. 

\subsection{Experimental Results}

The results are summarized in Table~\ref{tab:sygus_results}.
The \emph{Solved} column reports the number of benchmarks for which a {correct} solution was found within the 10-minute timeout.
\cvc successfully solved 137 out of 148 benchmarks. All 137 solutions were confirmed as correct by our post-synthesis verification.
The meaning of the other columns is the same as in the corresponding columns of Table~\ref{tab:ltl_results}.

\cvc solved 4 benchmarks that no LLM toolchain could solve.
\gpt solved 10 benchmarks that \cvc failed to solve.
We also observed that \gpt solves every benchmark that \qwen solved, while \qwen solved 6 benchmarks that \cvc failed to solve.
In total, 147 out of 148 benchmarks were solved by at least one method.

\begin{table*}
\centering
\begin{tabular}{l|c|ccc|cccc|cccc}
\hline
\textbf{Method} & \textbf{\# Solved} & \multicolumn{3}{|c|}{\textbf{Success Time}} & \multicolumn{4}{|c|}{\textbf{Success Iterations}} & \multicolumn{4}{c}{\textbf{Fail Iterations}} \\
\cline{3-5} \cline{6-9} \cline{10-13}
& within TO & Min & Max & Mean & Min & Max & Mean & Med & Min & Max & Mean & Med \\
\hline
\texttt{cvc5}    & 137       & 0.01 & 104.1 & 1.8   & --   & --   & --  & --  & --   & --   & --    & --     \\
\textsc{Qwen}    &  96       & 0.54 & 439.6 & 27.8  & 1    & 450  & 16.3& 1   & 15   & 921  & 432.5 & 466.5  \\
\textsc{GPT-5}   & 143       & 2.35 & 270.1 & 28.7  & --   & --   & --  & --  & --   & --   & --    & --     \\
\hline
\end{tabular}%

\caption{Summary statistics on 148 SyGuS benchmarks.\label{tab:sygus_results}}
\end{table*}

\subsubsection{LLM Response Failures}

The \qwen toolchain solved 96 benchmarks and reached the 10-minute timeout in the remaining $148 - 96 = 52$ benchmarks.
In all 52 failed cases, the LLM generated a final candidate that was not verified due to the timeout.
We used the verifier to check if these ``last-minute'' solutions were correct (note that they are not counted as ``solved'' in Table~\ref{tab:sygus_results}, even if verified to be correct).
Of these 52 candidates, 2 were actually correct, 46 were semantically incorrect (i.e., violating grammar or specification constraints), and 4 were syntactically incorrect (i.e., causing parsing errors in the verifier).

The \gpt toolchain failed to solve 5 out of 148 benchmarks.
Among these 5 failures, 3 were semantically incorrect, and 2 were caused by \gpt exceeding the token budget.

\subsubsection{\gpt: Tokens and Monetary Cost} 

For the \gpt toolchain, the average input length was 1,340 tokens per benchmark, while the average output length was 1,609 tokens. The total cost for all 148 queries was 2.62 USD.

\section{\TLA Distributed Protocol Synthesis}
\label{sec_tla_synth}

Given a {\em sketch} of a distributed protocol in \TLA~\cite{tla-lang}, the
synthesis problem is to {\em complete} the sketch such that the completed
protocol satisfies a given set of properties. A sketch is a \TLA protocol with
holes, along with grammars that define the set of expressions that can be used
to fill the holes. If the protocol is {\em parameterized}, e.g. by the
number of nodes participating in the protocol, we assume that the
parameters are finitized to specific values, which is standard practice~\cite{egolf2025tacas}.

\subsection{Experimental Setup}

\subsubsection{Symbolic Tool}

Our evaluation includes the state-of-the-art symbolic distributed protocol
synthesis tool \psem~\cite{egolf2025tacas}. \psem uses an {\em enumerative,
counterexample-guided} synthesis algorithm. Internally, it enumerates candidate
protocols, applies the TLC model checker~\cite{tlc-mc}, and halts if a
candidate protocol satisfies the properties. If not, it continues
enumerating candidates. Therefore, \psem cannot return a protocol that violates
the properties specified by the user~\cite{egolf2025tacas}. Because \psem
enumerates protocols from the language of the grammar~\cite{egolf2025tacas},
the synthesized protocol is guaranteed to conform to the grammar.

\subsubsection{LLM Toolchains}

We instantiate \gpt- and \qwen-based toolchains for this domain, and we use
the TLC model checker~\cite{tlc-mc} as the core verification engine. Because
grammar conformance is particularly important in this
domain~\cite{egolfFMCAD2024,egolf2025tacas}, our \qwen toolchain also uses a custom grammar verifier which checks
that the LLM responses conform to the provided grammar, before passing the
response to TLC. 
We use the same custom grammar verifier to check the solutions generated by \gpt.
Additionally, for this domain, our \qwen ILST loop maintains a cache mapping
LLM responses to model checking results. If an LLM repeats itself, we skip
model checking because we can infer that TLC has already rejected this
response.

\subsubsection{Prompt}

The prompt indicates which holes the LLM should
fill,
the actions where these holes appear,
the grammars for each hole,
the properties that the completed protocol must satisfy,
and the structure of the protocol in JSON format.
We show the prompt template for this domain
in
Appendix~\ref{app:tla_synth_prompt}.

\subsubsection{Benchmarks}

In~\cite{egolf2025tacas}, \psem was evaluated on a suite of 171 benchmarks, publicly
available at~\cite{polysemist-github}. We use these same benchmarks for
our evaluation. These benchmarks cover seven distributed protocols, including
two reconfigurable raft protocols. The benchmarks are to fill in missing
expressions within one or two actions of the protocol. These expressions are
either all pre-conditions for these actions, all post-conditions, or all pre-
and post-conditions. 

\subsubsection{Hardware and Parameters}

We run \psem on a machine with an AMD Ryzen 7 7840U CPU (3.3 GHz base clock)
and 14GB of RAM.
The \qwen toolchain ran on a machine with an Intel(R) Xeon(R) Platinum
8276 CPU (2.20GHz), 16GB RAM, plus an NVIDIA H200 GPU. We run GPT-5 using the
OpenAI API. For \gpt, we model check the responses obtained from the API on the
same machine used for \psem. The TLC model checker was run
with 8 workers in all toolchains and all toolchains had access to at least 8
CPU cores. There are no other sources of parallelism. We use the parameters
described in \S\ref{sec_common} for the \qwen and \gpt toolchains.

\subsection{Experimental Results}

The results are shown in Table~\ref{tab:tla-stats}. 
The {\em Solved} column shows for how many of the 171 benchmarks the corresponding method produces correct solutions, meaning solutions that pass both the model checker and the grammar verifier.
As shown in Table~\ref{tab:tla-stats}, the symbolic tool
\psem solves more benchmarks than \gptdists,
and \qwendists solves fewer.
There are benchmarks that are solved by the LLMs but not by
\psem, and vice versa. Together, \gptdists and \qwendists solve 17
benchmarks that \psem does not solve. Of these 17, \gptdists alone solves
11, \qwendists alone solves 1, and there are 5 benchmarks that both LLM
toolchains solve but \psem does not. There are 16 benchmarks that \psem
solves that neither LLM toolchain solves. There are 28 benchmarks that both
\gptdists and \psem solve, but \qwendists does not. There are 4 benchmarks
that both \qwendists and \psem solve, but \gptdists does not. There are 92
benchmarks that all three solve and 14 benchmarks that none of the three
solve. 

\begin{table*}
\centering
\small
\resizebox{\textwidth}{!}{%
\begin{tabular}{l|c|ccc|cccc|cccc}
\hline
\textbf{Method} & \textbf{\# Solved} & \multicolumn{3}{|c|}{\textbf{Success Time (s)}} & \multicolumn{4}{|c|}{\textbf{Success Iterations}} & \multicolumn{4}{c}{\textbf{Fail Iterations}} \\
\cline{3-5} \cline{6-9} \cline{10-13}
 & within TO & Min & Max & Mean & Min & Max & Mean & Med & Min & Max & Mean & Med \\
\hline
\psem & 140 & 0.70 & 583.06 & 70.94 & 1 & 219 & 20.7 & 10 & 7 & 272 & 97.3 & 99 \\
\qwendists & 102 & 2.03 & 587.07 & 35.03 & 1 & 254 & 12.2 & 1 & 76 & 717 & 228.9 & 202 \\
\gptdists & 136 & 17.55 & 275.35 & 102.22 & -- & -- & -- & -- & -- & -- & -- & -- \\
\hline
\end{tabular}%
}
\caption{Summary statistics on 171 TLA+ protocol synthesis benchmarks.\label{tab:tla-stats}}
\end{table*}

The meaning of the remaining columns of Table~\ref{tab:tla-stats} is the same as the corresponding columns in Table~\ref{tab:ltl_results}.
In the case of \psem we also report iteration statistics, which are available for this symbolic tool~\cite{egolf2025tacas}.

\subsubsection{Cactus Plots}

\begin{figure}
\centering
\begin{tikzpicture}
  \begin{axis}[
    width=12cm, height=6cm, xlabel={Time (s)}, ylabel={Number of Solved Instances}, grid=major, legend style={at={(0.7,0.02)}, anchor=south, legend columns=2, font=\small}, xmax=600.00,
  ]
    \addplot+[const plot, line width=1.5pt, solid, mark=None] coordinates {
      (0.70, 1)
      (1.37, 2)
      (2.11, 3)
      (2.72, 4)
      (2.83, 5)
      (2.83, 6)
      (2.83, 7)
      (2.84, 8)
      (2.91, 9)
      (2.92, 10)
      (2.93, 11)
      (3.03, 12)
      (3.03, 13)
      (3.34, 14)
      (3.51, 15)
      (3.52, 16)
      (3.54, 17)
      (3.65, 18)
      (3.85, 19)
      (3.91, 20)
      (4.03, 21)
      (4.24, 22)
      (4.26, 23)
      (4.37, 24)
      (4.44, 25)
      (4.45, 26)
      (4.49, 27)
      (4.57, 28)
      (4.70, 29)
      (4.76, 30)
      (4.95, 31)
      (5.05, 32)
      (5.08, 33)
      (5.11, 34)
      (5.13, 35)
      (5.19, 36)
      (5.25, 37)
      (5.53, 38)
      (5.72, 39)
      (5.73, 40)
      (6.09, 41)
      (6.17, 42)
      (6.28, 43)
      (6.43, 44)
      (6.53, 45)
      (6.57, 46)
      (6.89, 47)
      (7.05, 48)
      (7.07, 49)
      (7.11, 50)
      (7.13, 51)
      (7.18, 52)
      (7.19, 53)
      (7.28, 54)
      (7.31, 55)
      (7.31, 56)
      (7.32, 57)
      (7.45, 58)
      (7.66, 59)
      (7.76, 60)
      (7.78, 61)
      (7.79, 62)
      (7.88, 63)
      (7.90, 64)
      (8.00, 65)
      (8.07, 66)
      (8.51, 67)
      (8.81, 68)
      (8.95, 69)
      (9.82, 70)
      (9.92, 71)
      (9.98, 72)
      (10.74, 73)
      (11.52, 74)
      (11.57, 75)
      (11.67, 76)
      (11.95, 77)
      (12.03, 78)
      (12.45, 79)
      (12.50, 80)
      (12.82, 81)
      (12.87, 82)
      (13.70, 83)
      (13.85, 84)
      (14.02, 85)
      (14.09, 86)
      (14.17, 87)
      (14.67, 88)
      (17.94, 89)
      (26.16, 90)
      (26.67, 91)
      (30.93, 92)
      (35.02, 93)
      (37.48, 94)
      (39.26, 95)
      (41.73, 96)
      (58.93, 97)
      (62.98, 98)
      (64.59, 99)
      (68.29, 100)
      (69.62, 101)
      (69.72, 102)
      (73.16, 103)
      (73.48, 104)
      (77.18, 105)
      (79.02, 106)
      (80.71, 107)
      (84.12, 108)
      (86.68, 109)
      (101.97, 110)
      (104.56, 111)
      (115.67, 112)
      (118.39, 113)
      (118.75, 114)
      (123.48, 115)
      (125.49, 116)
      (128.54, 117)
      (133.86, 118)
      (143.85, 119)
      (144.49, 120)
      (154.33, 121)
      (154.70, 122)
      (176.88, 123)
      (185.20, 124)
      (189.21, 125)
      (199.65, 126)
      (215.34, 127)
      (247.85, 128)
      (253.45, 129)
      (306.19, 130)
      (339.21, 131)
      (343.36, 132)
      (405.24, 133)
      (450.00, 134)
      (462.98, 135)
      (487.77, 136)
      (523.05, 137)
      (535.21, 138)
      (552.85, 139)
      (583.06, 140)
      (584.06, 140)
      (600.00, 140)
    };
    \addlegendentry{\psem}

    \addplot+[const plot, line width=1.5pt, dash pattern=on 2pt off 2pt, mark=None] coordinates {
      (2.03, 1)
      (2.05, 2)
      (2.17, 3)
      (2.17, 4)
      (2.31, 5)
      (2.67, 6)
      (2.70, 7)
      (2.71, 8)
      (2.74, 9)
      (2.75, 10)
      (2.79, 11)
      (2.83, 12)
      (2.92, 13)
      (2.95, 14)
      (3.00, 15)
      (3.02, 16)
      (3.04, 17)
      (3.09, 18)
      (3.12, 19)
      (3.12, 20)
      (3.17, 21)
      (3.19, 22)
      (3.20, 23)
      (3.48, 24)
      (3.59, 25)
      (3.63, 26)
      (3.65, 27)
      (3.66, 28)
      (3.66, 29)
      (3.66, 30)
      (3.69, 31)
      (3.71, 32)
      (3.78, 33)
      (3.86, 34)
      (3.92, 35)
      (4.06, 36)
      (4.11, 37)
      (4.11, 38)
      (4.14, 39)
      (4.32, 40)
      (4.36, 41)
      (4.37, 42)
      (4.38, 43)
      (4.45, 44)
      (4.51, 45)
      (4.57, 46)
      (4.62, 47)
      (4.75, 48)
      (4.77, 49)
      (4.98, 50)
      (5.08, 51)
      (5.16, 52)
      (5.30, 53)
      (5.33, 54)
      (5.56, 55)
      (5.57, 56)
      (5.66, 57)
      (5.87, 58)
      (6.02, 59)
      (6.18, 60)
      (6.19, 61)
      (6.41, 62)
      (7.24, 63)
      (7.29, 64)
      (7.39, 65)
      (7.43, 66)
      (9.06, 67)
      (9.56, 68)
      (11.47, 69)
      (12.98, 70)
      (16.68, 71)
      (16.83, 72)
      (17.20, 73)
      (17.62, 74)
      (18.18, 75)
      (18.27, 76)
      (18.29, 77)
      (18.36, 78)
      (18.37, 79)
      (18.69, 80)
      (23.37, 81)
      (26.04, 82)
      (28.40, 83)
      (28.79, 84)
      (29.44, 85)
      (32.76, 86)
      (38.26, 87)
      (46.60, 88)
      (47.09, 89)
      (48.37, 90)
      (62.15, 91)
      (68.83, 92)
      (69.93, 93)
      (72.36, 94)
      (83.35, 95)
      (87.45, 96)
      (101.21, 97)
      (115.96, 98)
      (371.53, 99)
      (535.80, 100)
      (574.17, 101)
      (587.07, 102)
      (588.07, 102)
      (600.00, 102)
    };
    \addlegendentry{\qwendists}

    \addplot+[const plot, line width=1.5pt, dash pattern=on 1pt off 1pt, mark=None] coordinates {
      (17.55, 1)
      (18.78, 2)
      (23.23, 3)
      (32.24, 4)
      (32.36, 5)
      (34.52, 6)
      (40.37, 7)
      (40.63, 8)
      (43.12, 9)
      (43.37, 10)
      (43.61, 11)
      (43.87, 12)
      (44.06, 13)
      (45.11, 14)
      (45.12, 15)
      (46.18, 16)
      (46.32, 17)
      (47.07, 18)
      (47.14, 19)
      (48.20, 20)
      (50.88, 21)
      (51.14, 22)
      (53.26, 23)
      (53.87, 24)
      (55.69, 25)
      (55.69, 26)
      (56.29, 27)
      (57.46, 28)
      (57.50, 29)
      (57.80, 30)
      (58.08, 31)
      (58.65, 32)
      (59.75, 33)
      (62.37, 34)
      (62.99, 35)
      (63.20, 36)
      (63.23, 37)
      (67.60, 38)
      (69.52, 39)
      (70.22, 40)
      (70.71, 41)
      (71.21, 42)
      (72.88, 43)
      (73.13, 44)
      (73.43, 45)
      (73.47, 46)
      (73.94, 47)
      (74.71, 48)
      (74.72, 49)
      (74.97, 50)
      (75.96, 51)
      (77.88, 52)
      (79.23, 53)
      (79.28, 54)
      (80.71, 55)
      (81.94, 56)
      (82.11, 57)
      (82.16, 58)
      (83.90, 59)
      (84.50, 60)
      (84.98, 61)
      (85.15, 62)
      (86.21, 63)
      (86.35, 64)
      (88.78, 65)
      (89.00, 66)
      (89.33, 67)
      (90.39, 68)
      (90.59, 69)
      (90.95, 70)
      (93.18, 71)
      (94.30, 72)
      (94.77, 73)
      (95.17, 74)
      (97.72, 75)
      (98.98, 76)
      (99.30, 77)
      (100.57, 78)
      (100.88, 79)
      (101.23, 80)
      (102.48, 81)
      (103.53, 82)
      (106.53, 83)
      (107.00, 84)
      (108.41, 85)
      (108.96, 86)
      (110.87, 87)
      (111.27, 88)
      (112.26, 89)
      (113.00, 90)
      (116.83, 91)
      (118.72, 92)
      (119.47, 93)
      (120.61, 94)
      (121.02, 95)
      (121.52, 96)
      (122.64, 97)
      (124.24, 98)
      (125.16, 99)
      (125.29, 100)
      (127.68, 101)
      (129.15, 102)
      (133.08, 103)
      (134.76, 104)
      (138.82, 105)
      (139.11, 106)
      (141.04, 107)
      (141.43, 108)
      (142.42, 109)
      (142.52, 110)
      (144.26, 111)
      (147.03, 112)
      (149.24, 113)
      (154.15, 114)
      (156.65, 115)
      (165.20, 116)
      (165.53, 117)
      (165.85, 118)
      (168.55, 119)
      (169.95, 120)
      (171.92, 121)
      (174.95, 122)
      (175.57, 123)
      (177.36, 124)
      (178.82, 125)
      (178.84, 126)
      (179.21, 127)
      (184.63, 128)
      (187.14, 129)
      (197.49, 130)
      (209.24, 131)
      (226.57, 132)
      (238.23, 133)
      (260.89, 134)
      (262.11, 135)
      (275.35, 136)
      (276.35, 136)
      (600.00, 136)
    };
    \addlegendentry{\gptdists}
  \end{axis}
\end{tikzpicture}
\caption{If each benchmark is allowed a time budget (x-axis), the plot shows how many benchmarks can be solved (y-axis) within that time by each tool.}
\label{fig:tla-cactus}
\end{figure}
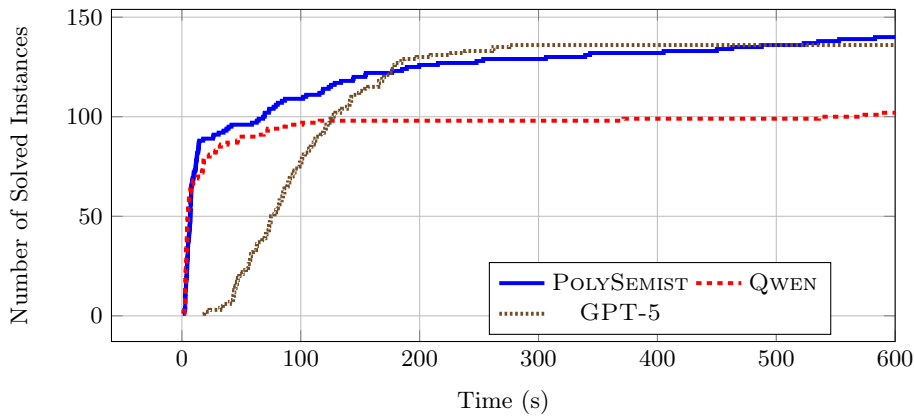

In addition to the  timing statistics in Table~\ref{tab:tla-stats},
Figure~\ref{fig:tla-cactus} shows cactus plots. The point ($t$, $c$) on each curve means that the corresponding method can solve $c$ benchmarks if each
benchmark is given a time budget of $t$ seconds (it does not mean that the tool
takes $t$ total seconds to solve $c$ benchmarks). The overall trend suggests
that \psem tends to solve the easier benchmarks faster than \gptdists. In
the long run, the two tools perform similarly.

\subsubsection{LLM Response Failures}

\gpt failed to solve 35 out of 171 benchmarks. In 33 of these failures the
model checker found counterexamples, and in one case \gpt exceeded the token
limit. In one case, \gpt produced a correct solution after more than 10
minutes, which we count as a timeout.

Because \qwen iterates, we received in total 16,824 responses from the LLM. If
we take the {\em sets} of distinct responses for each benchmark and sum their
cardinalities, we get 7,326 distinct responses. This means that \qwen repeats
itself (incorrectly) quite often. Of the 16,824 responses, 16,722 were failed
responses. In 11,628 of these cases, the model checker found
counterexamples. In one case, \qwen produced a correct solution after more than
10 minutes, which we count as a timeout. In 5,063 cases, the \qwen response did
not conform to the provided grammar (\gpt responses always conformed to the
provided grammar). In 30 cases \qwen did not even produce syntactically valid
JSON (\gpt always produced syntactically valid JSON).

\subsubsection{Relaxing Grammar Constraints}

We also evaluated the LLM toolchains with a relaxed verification engine that
does not check for grammar conformance but only applies TLC to check for semantic
correctness.
In the case of \qwen this means that in the ILST loop, only TLC is used as the verifier, and when a semantically valid solution is found the loop exits.
 Under this relaxed setting, we also omitted the grammar from the
prompt for both \qwen and \gpt. We found that both LLM toolchains solve more benchmarks in this
configuration. This result is not surprising, as the LLMs in this case are
solving an easier synthesis problem (without the grammar constraint). Under this relaxation, 
\qwen solves 133 benchmarks (compared to 102 in Table~\ref{tab:tla-stats}), while
\gpt solves 149 benchmarks (compared to 136 in Table~\ref{tab:tla-stats}, and compared to the 140 that \psem solves).

\subsubsection{\gpt: Tokens and Monetary Cost}

We use \gptdistww to refer to the \gpt toolchain where the grammar is not
checked (for the relaxation discussed above). For \gptdists, the average number
of input tokens per benchmark is
1953, and for \gptdistww it is 1220. The average number of output tokens per
benchmark is for \gptdists, and 5724 for \gptdistww. We note that on
average the synthesized solution in both cases is only about 70 tokens and the
majority of output tokens are the so-called ``reasoning tokens.'' \gptdistww is
given fewer input tokens because the grammar is not included in the prompt. We
note that \gptdistww produces more reasoning tokens; removing the grammar
causes the model to ``think'' more. 
Overall, the total cost to run all GPT-5 experiments reported for this domain
is about 20 USD.

\section{ACL2s Recursive Program Synthesis}
\label{sec_acl2s_synth}

The recursive program synthesis problem considered in this domain is to complete sketches
for one or more recursive functions such that the program composed of the completed functions satisfies a given set of correctness properties. 
As in \S\ref{sec_tla_synth}, a sketch is a program with holes, along with grammars that define the set of expressions that can be used to fill the holes. 
In this domain, the programs are written in the LISP-like ACL2s programming language~\cite{acl2s},
while sketches and grammars are expressed in a similar LISP-like format.
The correctness properties are first-order logic formulas.
We refer the reader to~\cite{egolf2026recursiveprogramsynthesissketches} for details.

\subsection{Experimental Setup}

\subsubsection{Symbolic Tool}

We use \cataclyst~\cite{egolf2026recursiveprogramsynthesissketches}, a symbolic
tool that synthesizes recursive programs from formal specifications. 
\cataclyst employs an enumerative, counterexample-guided synthesis algorithm.
\cataclyst internally checks candidate programs using the verification engine
discussed below. A consequence is that \cataclyst can only output programs that
pass the verification engine. It also cannot output syntactically invalid
programs.

\subsubsection{Verifier and Correctness Guarantees}

Verification of synthesized programs in this domain amounts to first-order
logic validity checking, and is therefore undecidable in general. We therefore
follow the approach taken in~\cite{egolf2026recursiveprogramsynthesissketches},
and use as the verification engine of our toolchains the counterexample
generator of the ACL2s theorem prover~\cite{acl2s,acl2s-cgen}. A consequence of
this is that \cataclyst can, in principle, return a program that passes the
counterexample generator but is not actually correct. The authors
of~\cite{egolf2026recursiveprogramsynthesissketches} use the ACL2s theorem
prover to perform post-hoc manual verification of all programs synthesized
in~\cite{egolf2026recursiveprogramsynthesissketches} and found that all
synthesized programs were indeed correct. We therefore take for granted that,
at least for the benchmarks
in~\cite{egolf2026recursiveprogramsynthesissketches} (which we also use
here), passing counterexample generation is a good proxy for full-blown
verification.

\subsubsection{LLM Toolchains}

As for the prior domains, the LLM toolchains have the same overall structure as
in Figure~\ref{fig:llm-method}. We instantiate two LLM toolchains: one using
\qwen and one using \gpt. Both toolchains are relaxed, in the sense that they check the correctness properties but 
do not check that proposed programs conform to the provided sketches or
grammars. In this domain, the sketches and grammars are primarily useful for
guiding \cataclyst's enumerative search. It is not critical that the
synthesized programs conform to the grammar, so long as the programs are
syntactically valid. In the interest of fairness though, we give the LLM
toolchains the sketches and grammars as input. Here, we also cache LLM responses and
skip verification if the LLM repeats itself.

\subsubsection{Prompts}

For this domain, the prompt template gives exposes: (1) the function signatures
of the functions under synthesis, (2) the list of primitive functions that can
be used in the synthesized functions, (3) the terminals (constants and function
arguments) that can be used in the synthesized functions, (4) the list of
properties and input-output examples that the synthesized functions must
satisfy, (5) auxiliary information about data types and functions that may be
useful context for the LLM, and (6) the sketch for each function and the
grammar for filling holes in the sketches. The full prompt template that we use
is shown in Appendix~\ref{app:acl2s_synth_prompt}.

\subsubsection{Benchmarks}

We use the 42 benchmarks used to evaluate \cataclyst
in~\cite{egolf2026recursiveprogramsynthesissketches}. These benchmarks include
programs for manipulating lists, trees, and other inductive data structures. Some
benchmarks involve mutually recursive functions, and some benchmarks involve
properties with existential quantifiers.

\subsubsection{Hardware and Parameters}

We ran \cataclyst on a machine with an AMD Ryzen 7 7840U CPU (3.3 GHz base
clock) and 14GB of RAM. The \qwen hardware configuration is identical to
that described in \S\ref{sec_tla_synth} for the \TLA synthesis domain, except
that only one CPU core is allocated. The \gpt toolchain ran verification on the
same machine used for \cataclyst and uses the same LLM parameters as described
in \S\ref{sec_common}. Neither the symbolic tool nor the verifier are
configured to use parallelism.

\subsection{Experimental Results}

Our results are summarized in Table~\ref{tab:acl2s-stats}. All methods in this domain use a 15 minute
timeout rather than the 10 minute timeout used in prior domains. 
The {\em Solved} column reports solutions that pass the ACL2s counterexample generator.
The meaning of the other columns is the same as in the corresponding tables of prior domains.
As shown in Table~\ref{tab:acl2s-stats}, \qwenrec solved
39/42 benchmarks and both \gptrec and \cataclyst solved 41/42 benchmarks.
\qwenrec failed on the following three benchmarks: ``pairs,''  ``split-on,''
and ``suffixb''~\cite{egolf2026recursiveprogramsynthesissketches}. \gptrec failed only on ``suffixb.'' \cataclyst failed only on
``ternary-tree-eq.'' So both LLM toolchains struggled with ``suffixb,'' which
\cataclyst was able to solve, but both LLM toolchains succeeded on
``ternary-tree-eq,'' which \cataclyst could not solve. 

\begin{table*}
\centering
\small
\resizebox{\textwidth}{!}{%
\begin{tabular}{l|c|ccc|cccc|cccc}
\hline
\textbf{Method} & \textbf{\# Solved} & \multicolumn{3}{|c|}{\textbf{Success Time (s)}} & \multicolumn{4}{|c|}{\textbf{Success Iterations}} & \multicolumn{4}{c}{\textbf{Fail Iterations}} \\
\cline{3-5} \cline{6-9} \cline{10-13}
 & within TO & Min & Max & Mean & Min & Max & Mean & Med & Min & Max & Mean & Med \\
\hline
\cataclyst & 41 & 12.60 & 110.23 & 39.11 & 4 & 2552 & 495.6 & 173 & 4795 & 4795 & 4795 & 4795 \\
\qwenrec & 39 & 3.23 & 30.42 & 9.03 & 1 & 6 & 1.3 & 1 & 179 & 516 & 369.7 & 414 \\
\gptrec & 41 & 17.33 & 124.16 & 48.33 & -- & -- & -- & -- & -- & -- & -- & -- \\
\hline
\end{tabular}%
}
\caption{Summary statistics on 42 ACL2s recursive program synthesis benchmarks.}
\label{tab:acl2s-stats}
\end{table*}

\subsubsection{Iterations and Failures}

As shown in Table~\ref{tab:acl2s-stats}, in this domain,
when \qwenrec solves a benchmark, it tends to do so using just a few iterations (and corresponding LLM
queries). On the other hand, when \cataclyst solves a benchmark, it often
enumerates hundreds of candidate solutions. When \qwenrec does not solve a
benchmark, it does so after proposing hundreds of incorrect solutions: 179,
414, and 516 proposals for the three unsolved benchmarks. In the single
case where \cataclyst fails, it does so after enumerating 4795 candidate
solutions.

We count also the number of {\em distinct} solutions proposed by \qwenrec in
the three failed benchmarks: 48/179, 7/414, 2/516. I.e., in one of the failed
tasks \qwenrec was asked 179 times for a solution, and in all but 48 cases
it proposed a solution that it had proposed before. In general, we obtained a
total of 1159 responses from \qwenrec across all benchmarks. Only 165 of those responses were 
distinct, so \qwen repeated itself many times.

Of the 1159 responses we obtained from \qwenrec, 1120 were failed responses.
501 of these failures were caused because the counterexample generator
disproved correctness of the proposed solution. 619 failures were due to
syntactically invalid responses from the LLM, e.g., mismatched parentheses.
In the one case where \gpt failed, it was due to a counterexample being found.

\subsubsection{\gpt: Tokens and Monetary Cost}

For \gptrec, the average number of input tokens per benchmark is 681, the
solution is composed of 87 output tokens on average, and GPT-5 outputs an
average of 1761 reasoning tokens per benchmark. 
The total monetary cost for all 42 benchmarks is about 0.80 USD.

\section{Related Work}
\label{sec:related}

In the LTL reactive synthesis domain, the works~\cite{SchmittRabeFinkbeiner2021,schmitt2023neuralCircuitSynthesisPreTrained,CoslerHahnNeuroSynt24} explore neural approaches and evaluate them on SYNTCOMP benchmarks.
All these works train or fine-tune 
\emph{transformer} models for synthesizing AIGER circuits, 
whereas we evaluate off-the-shelf models without any additional training,
and we consider both AIGER and SMV outputs.

In the SyGuS domain, Li et al.~\cite{liCAV2024SyGuSwithLLMs} propose a hybrid approach that integrates LLM calls into a weighted probabilistic enumerative search.
Their evaluation shows that \cvc outperforms GPT-3.5 alone, while their hybrid approach outperforms both.
Our work compares \cvc with the more recent \gpt; in our experiments, \gpt outperforms \cvc.
We also compare with the 32-billion parameter Qwen-2.5-Coder, whereas~\cite{liCAV2024SyGuSwithLLMs} only compare with GPT-3.5.
Finally, we use a different set of benchmarks than those used in~\cite{liCAV2024SyGuSwithLLMs}. 

To our knowledge,~\cite{egolfFMCAD2024,egolf2025tacas} are the only works
that synthesize protocols in \TLA, and they do so with
symbolic tools with no mention of LLMs. The GenAI-accelerated TLA+
challenge~\cite{genaiTLAChallenge2025} called ``for submissions
showcasing creative and technically impressive work at the intersection of
\TLA, formal methods, and AI-assisted development,'' and announced winners in
2025. The first place winner Specula takes as input program source code (e.g. Rust) and
outputs \TLA. The third place winner by Gregory Terzian takes as input
\TLA and produces Rust code. These tools solve \TLA code generation problems using LLMs, but these problems are different from the protocol synthesis problem we consider in this paper. 

\cite{smyth,escher,para,myth,lambda-sq} synthesize recursive programs from
input-output examples.
\cite{contata,burst,synquid,leon,egolf2026recursiveprogramsynthesissketches}
synthesize recursive programs from properties. None of these works compare LLMs
with symbolic tools.

There are other synthesis domains that fall outside the scope of this paper.
LLMs are used to generate 
loop invariants in~\cite{chakraborty2023rankingLLMLoopInvariants,kamath2023findingIndLoopInvariantsLLM}, 
class invariants in~\cite{DBLP:conf/saiv/SunACTBDQL25},
spreadsheet formulas in~\cite{joshi2024flameExcelFormulasSynthesisLM},
and to perform C program repair in~\cite{tihanyi2024newerasoftwaresecurity}.
These works address different synthesis problems.

Several of the works cited above can be considered ``hybrid'' or ``neurosymbolic'' in the sense that they combine neural networks and symbolic tools (including verifiers, but in more sophisticated ways than our {\em iterate-LLM-until-solution-or-timeout} approach).  
Neurosymbolic synthesis~\cite{chaudhuri2025neurosymbolic} is an active research direction but beyond the scope of our current evaluation.

Several works propose benchmarks for LLM code generation in various domains, including general programming tasks~\cite{chen2021humanEval,zhuo2024bigcodebench,openai2025competitiveprogramminglargereasoning,qiu2025how,liu2024evalPerf,huang2024effiBench}, 
as well as code generation tasks more closely related to this paper.
The works in~\cite{ye2025verina,thakur2025cleverBenchmark,bursuc2025vericodingBenchmark,loughridge2025dafnybench} propose benchmarks for verification-oriented languages such as Lean~\cite{moura2021lean4}, Dafny~\cite{leino2010dafny}, or Verus~\cite{lattuada2023verus}.
LLM code generation benchmarks for Solidity smart contracts are proposed in~\cite{peng2025SolEvalSmartContractBenchmark},
and for the domain of hardware  in~\cite{kang2025FVEvalHardwareVerificationBencmhmark,abdelatty2025pluto,jin2025RealBenchVerilogBenchmark,liu2023VerilogEvalBenchmark,lu2024RTLLMHardwareBenchmark}.
Several of these works evaluate LLM-generated code with respect to correctness or efficiency, and several employ verification tools for checking correctness.
However, the focus of these works is not the comparison of LLMs with symbolic tools, 
and their domains are different from the ones examined in our work.

\section{Conclusions}
\label{sec_conclusion}

We compared symbolic synthesis tools with LLM toolchains, across several synthesis domains.
Although the results vary by domain, the following observations can be made
(c.f. Tables~\ref{tab:ltl_results},\ref{tab:sygus_results},\ref{tab:tla-stats},\ref{tab:acl2s-stats}
and the corresponding sections):

\noindent
{\bf(1)} The symbolic tools solve more benchmarks than
the \qwen toolchains, in all domains.
The difference is dramatic in the LTL synthesis domain.

\noindent
{\bf(2)} \ltlsynt solves significantly more benchmarks than the \gpt toolchains in
the LTL synthesis domain. \gpt solves almost the same number of benchmarks as
the corresponding symbolic tool in the other three domains.

\noindent
{\bf(3)} There are generally benchmarks that are solved by the LLM but not by the symbolic tool, and vice versa.
From a practical perspective, this suggests using a portfolio approach which runs both symbolic and LLM-based toolchains in parallel.

\noindent
{\bf(4)} The symbolic tools are faster than the \gpt toolchains, in all domains, despite the fact that \gpt runs on more powerful (and proprietary) hardware.
In the LTL synthesis and SyGuS domains, the difference is two and one orders of magnitude, respectively.

\noindent
{\bf(5)} Even though this may not be immediately apparent in the tables above, the
symbolic tools are also faster than the \qwen toolchains, in all domains
except recursive program synthesis.
To see this, note first that the tables show timing statistics
about {\em successful} runs only. To get the full picture, the complete
execution times must be inferred. For instance, consider
Table~\ref{tab:tla-stats}. \qwen solves 102 benchmarks with an average of 35.03
secs per benchmark. But \qwen also times out on 69 benchmarks, each after 600
secs. In total, \qwen executed for $35.03\cdot 102 + 600\cdot 69 = 44973$ secs
and managed to solve 102 benchmarks in that time. In contrast, \psem executed
in total for $70.94\cdot 140 + 600\cdot 31 = 28532$ secs and managed to solve
140 benchmarks in that time. The same calculation can be done for
Table~\ref{tab:acl2s-stats}.  \qwen runs for 2152.17 seconds, which
is slightly less than \cataclyst's 2203.51 seconds; 
but \qwen is run on significantly more powerful hardware (a GPU)
compared to \cataclyst (a CPU), and \qwen solves fewer benchmarks (39 vs
41).

\noindent
{\bf(6)} \qwen toolchains often repeat the same incorrect proposal many times (c.f. \S\ref{sec_tla_synth} and \S\ref{sec_acl2s_synth}).
An advantage of symbolic tools that use an
enumerative synthesis algorithm is that they can be designed so that they never repeat
the same candidate.
In our \qwen toolchains,
we sample many times, but we do not show the LLM past proposals and ask it for
something new. It is possible that doing so would reduce the number of repeated
proposals, but LLMs offer no such guarantees. Furthermore, concatenating past
proposals may become infeasible as the number of proposals becomes large,
due to token and context window limits.

\noindent
{\bf(7)} Except for the LTL domain, \gpt was quite good at meeting the syntactic
expectations that we set for it (well-formed code, grammar/sketch conformance).
\qwen struggled  with meeting these expectations. Even in the recursive
program synthesis domain, where it ultimately solved many benchmarks, more
than half of its failures were due to syntactic issues. Symbolic tools are
designed to always meet syntactic expectations, which is an advantage.


\begin{credits}
\subsubsection{\ackname}
This material is partly supported by the National Science
Foundation under Graduate Research Fellowship Grant \#1938052.
This work has also been partly supported by NSF's FMitF (Formal Methods in the Field) program, under awards 2319500 and 2525087.
Any opinion,
findings, and conclusions or recommendations expressed in this material are
those of the authors(s) and do not necessarily reflect the views of the National
Science Foundation.
\end{credits}

%
%
%
\bibliographystyle{splncs04}
\bibliography{bib}

@inproceedings{Church57,
author={Alonzo Church},
title={Applications of recursive arithmetic to the problem of circuit synthesis},
booktitle={Summaries of the Summer Institute of Symbolic Logic},
volume=1,
pages={3-50},
year=1957,
}

@inproceedings{PnueliLTL77,
 author = {Pnueli, Amir},
 title = {The Temporal Logic of Programs},
 booktitle = {Proceedings of the 18th Annual Symposium on Foundations of Computer Science},
 series = {SFCS '77},
 year = 1977,
 pages = {46--57},
}

@article{MannaWaldinger1980,
 author = {Manna, Zohar and Waldinger, Richard},
 title = {A Deductive Approach to Program Synthesis},
 journal = {ACM Trans. Program. Lang. Syst.},
 volume = {2},
 number = {1},
 month = jan,
 year = 1980,
 issn = {0164-0925},
 pages = {90--121},
 numpages = {32},
 doi = {10.1145/357084.357090},
 acmid = {357090},
 publisher = {ACM},
}

@inproceedings{PnueliRosner89,
  author={Amir Pnueli and Roni Rosner},
  title={On the synthesis of a reactive module},
  booktitle={{ACM} Symp. POPL},
  year=1989,
}

@inproceedings{Finkbeiner2016NATO,
  author = {Bernd Finkbeiner},
  title = {Synthesis of Reactive Systems},
  editor    = {Javier Esparza and Orna Grumberg and Salomon Sickert},
  booktitle = {Dependable Software Systems Engineering},
  series    = {{NATO} Science for Peace and Security Series, {D:} Information and Communication Security},
  volume    = {45},
  pages = {72--98},
  publisher = {{IOS} Press},
  year      = 2016,
}

@article{GulwaniPolozovSingh2017,
year = 2017,
volume = {4},
journal = {Foundations and Trends in Programming Languages},
title = {Program Synthesis},
doi = {10.1561/2500000010},
issn = {2325-1107},
number = {1-2},
pages = {1-119},
author = {Sumit Gulwani and Oleksandr Polozov and Rishabh Singh}
}

@inproceedings{Solar-LezamaPLDI2005,
 author = {Solar-Lezama, Armando and Rabbah, Rodric and Bod\'{\i}k, Rastislav and Ebcio\u{g}lu, Kemal},
 title = {Programming by Sketching for Bit-streaming Programs},
 booktitle = {Proceedings of the 2005 ACM SIGPLAN Conference on Programming Language Design and Implementation},
 series = {PLDI '05},
 year = {2005},
 isbn = {1-59593-056-6},
 pages = {281--294},
 numpages = {14},
 url = {http://doi.acm.org/10.1145/1065010.1065045},
 acmid = {1065045},
 publisher = {ACM},
}

@article{ArmandoSTTT2013,
author = {Solar-Lezama, Armando},
title = {Program sketching},
year = 2013,
publisher = {Springer},
volume = {15},
number = {5-6},
issn = {1433-2779},
url = {https://doi.org/10.1007/s10009-012-0249-7},
doi = {10.1007/s10009-012-0249-7},
journal = {Int. J. Softw. Tools Technol. Transf.},
month = {oct},
pages = {475-495},
numpages = {21}
}

@inproceedings{Sygus13,
  author    = {R. Alur and
               R. Bod\'{\i}k and
               G. Juniwal and
               M.M.K. Martin and
               M. Raghothaman and
               S.A. Seshia and
               R. Singh and
               A. Solar-Lezama and
               E. Torlak and
               A. Udupa},
  title     = {Syntax-guided synthesis},
  booktitle = {Formal Methods in Computer-Aided Design, FMCAD},
  year      = 2013,
  pages     = {1-17},
}

@misc{tihanyi2024newerasoftwaresecurity,
      title={A New Era in Software Security: Towards Self-Healing Software via Large Language Models and Formal Verification}, 
      author={Norbert Tihanyi and Ridhi Jain and Yiannis Charalambous and Mohamed Amine Ferrag and Youcheng Sun and Lucas C. Cordeiro},
      year={2024},
      eprint={2305.14752},
      archivePrefix={arXiv},
      primaryClass={cs.SE},
      url={https://arxiv.org/abs/2305.14752}, 
}

@inproceedings{SchmittRabeFinkbeiner2021,
 author = {Schmitt, Frederik and Hahn, Christopher and Rabe, Markus N and Finkbeiner, Bernd},
 booktitle = {Advances in Neural Information Processing Systems},
 editor = {M. Ranzato and A. Beygelzimer and Y. Dauphin and P.S. Liang and J. Wortman Vaughan},
 pages = {15408--15420},
 publisher = {Curran Associates, Inc.},
 title = {Neural Circuit Synthesis from Specification Patterns},
 url = {https://proceedings.neurips.cc/paper_files/paper/2021/file/8230bea7d54bcdf99cdfe85cb07313d5-Paper.pdf},
 volume = {34},
 year = {2021}
}

@InProceedings{CoslerHahnNeuroSynt24,
author="Cosler, Matthias
and Hahn, Christopher
and Omar, Ayham
and Schmitt, Frederik",
editor="Finkbeiner, Bernd
and Kov{\'a}cs, Laura",
title="NeuroSynt: A Neuro-symbolic Portfolio Solver for Reactive Synthesis",
booktitle="Tools and Algorithms for the Construction and Analysis of Systems",
year="2024",
publisher="Springer Nature Switzerland",
address="Cham",
pages="45--67",
isbn="978-3-031-57256-2"
}

@inproceedings{DBLP:conf/saiv/SunACTBDQL25,
  author       = {Chuyue Sun and
                  Viraj Agashe and
                  Saikat Chakraborty and
                  Jubi Taneja and
                  Clark W. Barrett and
                  David L. Dill and
                  Xiaokang Qiu and
                  Shuvendu K. Lahiri},
  editor       = {Mirco Giacobbe and
                  Anna Lukina},
  title        = {{ClassInvGen: Class Invariant Synthesis Using Large Language Models}},
  booktitle    = {{AI} Verification - Second International Symposium, {SAIV} 2025, Zagreb,
                  Croatia, July 21-22, 2025, Proceedings},
  series       = {Lecture Notes in Computer Science},
  volume       = {15947},
  pages        = {64--96},
  publisher    = {Springer},
  year         = {2025},
  url          = {https://doi.org/10.1007/978-3-031-99991-8\_4},
}

@inproceedings{egolfFMCAD2024,
      title={{Efficient Synthesis of Symbolic Distributed Protocols by Sketching}},
      author={Derek Egolf and William Schultz and Stavros Tripakis},
  booktitle={FMCAD 2024: Formal Methods in Computer-Aided Design},
      year=2024,
}

@inproceedings{egolf2025tacas,
      title={{Accelerating Protocol Synthesis and Detecting Unrealizability with Interpretation Reduction}},
      author={Derek Egolf and Stavros Tripakis},
        booktitle={31st International Conference on Tools and Algorithms for the Construction and Analysis of Systems (TACAS)},
      year=2025,
}

@misc{egolf2026recursiveprogramsynthesissketches,
      title={Recursive Program Synthesis from Sketches and Mixed-Quantifier Properties},
      author={Derek Egolf and Stavros Tripakis},
      year=2026,
      eprint={2601.04045},
      archivePrefix={arXiv},
      primaryClass={cs.LO},
      url={https://arxiv.org/abs/2601.04045},
}

@inproceedings{liCAV2024SyGuSwithLLMs,
  title={Guiding enumerative program synthesis with large language models},
  author={Li, Yixuan and Parsert, Julian and Polgreen, Elizabeth},
  booktitle={International Conference on Computer Aided Verification},
  pages={280--301},
  year={2024},
  organization={Springer}
}

@inproceedings{schmitt2023neuralCircuitSynthesisPreTrained,
  title={Neural circuit synthesis with pre-trained language models},
  author={Schmitt, Frederik and Cosler, Matthias and Finkbeiner, Bernd},
  booktitle={First International Workshop on Deep Learning-aided Verification},
  year={2023}
}

@article{ye2025verina,
  title={{Verina: Benchmarking Verifiable Code Generation}},
  author={Ye, Zhe and Yan, Zhengxu and He, Jingxuan and Kasriel, Timothe and Yang, Kaiyu and Song, Dawn},
  journal={arXiv preprint arXiv:2505.23135},
  year={2025}
}

@article{thakur2025cleverBenchmark,
  title={{CLEVER}: A Curated Benchmark for Formally Verified Code Generation},
  author={Thakur, Amitayush and Lee, Jasper and Tsoukalas, George and Sistla, Meghana and Zhao, Matthew and Zetzsche, Stefan and Durrett, Greg and Yue, Yisong and Chaudhuri, Swarat},
  journal={arXiv preprint arXiv:2505.13938},
  year={2025}
}

@article{bursuc2025vericodingBenchmark,
  title={A benchmark for vericoding: Formally verified program synthesis},
  author={Bursuc, Sergiu and Ehrenborg, Theodore and Lin, Shaowei and Astefanoaei, Lacramioara and Chiosa, Ionel Emilian and Kukovec, Jure and Singh, Alok and Butterley, Oliver and Bizid, Adem and Dougherty, Quinn and others},
  journal={arXiv preprint arXiv:2509.22908},
  year={2025}
}

@inproceedings{peng2025SolEvalSmartContractBenchmark,
  title={{SolEval: Benchmarking Large Language Models for Repository-level Solidity Smart Contract Generation}},
  author={Peng, Zhiyuan and Yin, Xin and Qian, Rui and Lin, Peiqin and Liu, Yongkang and Zhang, Hao and Ying, Chenhao and Luo, Yuan},
  booktitle={Proceedings of the 2025 Conference on Empirical Methods in Natural Language Processing},
  pages={4388--4411},
  year={2025}
}

@inproceedings{kang2025FVEvalHardwareVerificationBencmhmark,
  title={{FVEval}: Understanding language model capabilities in formal verification of digital hardware},
  author={Kang, Minwoo and Liu, Mingjie and Hamad, Ghaith Bany and Suhaib, Syed M and Ren, Haoxing},
  booktitle={2025 Design, Automation \& Test in Europe Conference (DATE)},
  pages={1--6},
  year={2025},
  organization={IEEE}
}

@article{jin2025RealBenchVerilogBenchmark,
  title={{RealBench}: Benchmarking {Verilog} generation models with real-world {IP} designs},
  author={Jin, Pengwei and Huang, Di and Li, Chongxiao and Cheng, Shuyao and Zhao, Yang and Zheng, Xinyao and Zhu, Jiaguo and Xing, Shuyi and Dou, Bohan and Zhang, Rui and others},
  journal={arXiv preprint arXiv:2507.16200},
  year={2025}
}

@inproceedings{liu2023VerilogEvalBenchmark,
  title={{VerilogEval}: Evaluating large language models for {Verilog} code generation},
  author={Liu, Mingjie and Pinckney, Nathaniel and Khailany, Brucek and Ren, Haoxing},
  booktitle={2023 IEEE/ACM International Conference on Computer Aided Design (ICCAD)},
  pages={1--8},
  year={2023},
  organization={IEEE}
}

@inproceedings{lu2024RTLLMHardwareBenchmark,
  title={{RTLLM}: An open-source benchmark for design {RTL} generation with large language model},
  author={Lu, Yao and Liu, Shang and Zhang, Qijun and Xie, Zhiyao},
  booktitle={2024 29th Asia and South Pacific Design Automation Conference (ASP-DAC)},
  pages={722--727},
  year={2024},
  organization={IEEE}
}

@article{loughridge2025dafnybench,
title={{DafnyBench: A Benchmark for Formal Software Verification}},
author={Chloe R Loughridge and Qinyi Sun and Seth Ahrenbach and Federico Cassano and Chuyue Sun and Ying Sheng and Anish Mudide and Md Rakib Hossain Misu and Nada Amin and Max Tegmark},
journal={Transactions on Machine Learning Research},
issn={2835-8856},
year={2025},
url={https://openreview.net/forum?id=yBgTVWccIx},
note={}
}

@inproceedings{joshi2024flameExcelFormulasSynthesisLM,
  title={Flame: A small language model for spreadsheet formulas},
  author={Joshi, Harshit and Ebenezer, Abishai and Sanchez, Jos{\'e} Cambronero and Gulwani, Sumit and Kanade, Aditya and Le, Vu and Radi{\v{c}}ek, Ivan and Verbruggen, Gust},
  booktitle={Proceedings of the AAAI Conference on Artificial Intelligence},
  volume={38},
  number={12},
  pages={12995--13003},
  year={2024}
}

@article{chaudhuri2025neurosymbolic,
  title={Neurosymbolic program synthesis},
  author={Chaudhuri, Swarat},
  journal={Handbook on Neurosymbolic AI and Knowledge Graphs},
  volume={400},
  pages={532--549},
  year={2025},
  publisher={IOS Press}
}

@article{renkin2022ltlsynt,
  title={Dissecting ltlsynt},
  author={Renkin, Florian and Schlehuber-Caissier, Philipp and Duret-Lutz, Alexandre and Pommellet, Adrien},
  journal={Formal Methods in System Design},
  volume={61},
  number={2},
  pages={248--289},
  year={2022},
  publisher={Springer}
}

@inproceedings{chakraborty2023rankingLLMLoopInvariants,
  title={Ranking {LLM}-generated loop invariants for program verification},
  author={Chakraborty, Saikat and Lahiri, Shuvendu and Fakhoury, Sarah and Lal, Akash and Musuvathi, Madanlal and Rastogi, Aseem and Senthilnathan, Aditya and Sharma, Rahul and Swamy, Nikhil},
  booktitle={Findings of the Association for Computational Linguistics: EMNLP 2023},
  pages={9164--9175},
  year={2023}
}

@article{kamath2023findingIndLoopInvariantsLLM,
  title={Finding inductive loop invariants using large language models},
  author={Kamath, Adharsh and Senthilnathan, Aditya and Chakraborty, Saikat and Deligiannis, Pantazis and Lahiri, Shuvendu K and Lal, Akash and Rastogi, Aseem and Roy, Subhajit and Sharma, Rahul},
  journal={arXiv preprint arXiv:2311.07948},
  year={2023}
}

@article{abdelatty2025pluto,
  title={Pluto: A Benchmark for Evaluating Efficiency of {LLM}-generated Hardware Code},
  author={Abdelatty, Manar and Nouh, Maryam and Rosenstein, Jacob K and Reda, Sherief},
  journal={arXiv preprint arXiv:2510.14756},
  year={2025}
}

@misc{openai2025competitiveprogramminglargereasoning,
      title={Competitive Programming with Large Reasoning Models}, 
      author={OpenAI and : and Ahmed El-Kishky and Alexander Wei and Andre Saraiva and Borys Minaiev and Daniel Selsam and David Dohan and Francis Song and Hunter Lightman and Ignasi Clavera and Jakub Pachocki and Jerry Tworek and Lorenz Kuhn and Lukasz Kaiser and Mark Chen and Max Schwarzer and Mostafa Rohaninejad and Nat McAleese and o3 contributors and Oleg Mürk and Rhythm Garg and Rui Shu and Szymon Sidor and Vineet Kosaraju and Wenda Zhou},
      year={2025},
      eprint={2502.06807},
      archivePrefix={arXiv},
      primaryClass={cs.LG},
      url={https://arxiv.org/abs/2502.06807}, 
}

@article{chen2021humanEval,
  author       = {Mark Chen and
                  Jerry Tworek and
                  Heewoo Jun and
                  Qiming Yuan and
                  Henrique Pond{\'{e}} de Oliveira Pinto and
                  Jared Kaplan and
                  Harri Edwards and
                  Yuri Burda and
                  Nicholas Joseph and
                  Greg Brockman and
                  Alex Ray and
                  Raul Puri and
                  Gretchen Krueger and
                  Michael Petrov and
                  Heidy Khlaaf and
                  Girish Sastry and
                  Pamela Mishkin and
                  Brooke Chan and
                  Scott Gray and
                  Nick Ryder and
                  Mikhail Pavlov and
                  Alethea Power and
                  Lukasz Kaiser and
                  Mohammad Bavarian and
                  Clemens Winter and
                  Philippe Tillet and
                  Felipe Petroski Such and
                  Dave Cummings and
                  Matthias Plappert and
                  Fotios Chantzis and
                  Elizabeth Barnes and
                  Ariel Herbert{-}Voss and
                  William Hebgen Guss and
                  Alex Nichol and
                  Alex Paino and
                  Nikolas Tezak and
                  Jie Tang and
                  Igor Babuschkin and
                  Suchir Balaji and
                  Shantanu Jain and
                  William Saunders and
                  Christopher Hesse and
                  Andrew N. Carr and
                  Jan Leike and
                  Joshua Achiam and
                  Vedant Misra and
                  Evan Morikawa and
                  Alec Radford and
                  Matthew Knight and
                  Miles Brundage and
                  Mira Murati and
                  Katie Mayer and
                  Peter Welinder and
                  Bob McGrew and
                  Dario Amodei and
                  Sam McCandlish and
                  Ilya Sutskever and
                  Wojciech Zaremba},
  title        = {Evaluating Large Language Models Trained on Code},
  journal      = {CoRR},
  volume       = {abs/2107.03374},
  year         = {2021},
  url          = {https://arxiv.org/abs/2107.03374},
  eprinttype    = {arXiv},
  eprint       = {2107.03374},
}

@article{zhuo2024bigcodebench,
  title={{BigCodeBench}: Benchmarking code generation with diverse function calls and complex instructions},
  author={Zhuo, Terry Yue and Vu, Minh Chien and Chim, Jenny and Hu, Han and Yu, Wenhao and Widyasari, Ratnadira and Yusuf, Imam Nur Bani and Zhan, Haolan and He, Junda and Paul, Indraneil and others},
  journal={arXiv preprint arXiv:2406.15877},
  year={2024}
}

@inproceedings{qiu2025how,
title={{How efficient is {LLM}-generated code? A rigorous \& high-standard benchmark}},
author={Ruizhong Qiu and Weiliang Will Zeng and James Ezick and Christopher Lott and Hanghang Tong},
booktitle={The Thirteenth International Conference on Learning Representations},
year={2025},
url={https://openreview.net/forum?id=suz4utPr9Y}
}

@article{liu2024evalPerf,
      title={{Evaluating Language Models for Efficient Code Generation}}, 
  author={Liu, Jiawei and Xie, Songrun and Wang, Junhao and Wei, Yuxiang and Ding, Yifeng and Zhang, Lingming},
  journal={arXiv preprint arXiv:2408.06450},
  year={2024}
}

@article{huang2024effiBench,
  title={{EffiBench: Benchmarking the efficiency of automatically generated code}},
  author={Huang, Dong and Qing, Yuhao and Shang, Weiyi and Cui, Heming and Zhang, Jie M},
  journal={Advances in Neural Information Processing Systems},
  volume={37},
  pages={11506--11544},
  year={2024}
}

@inproceedings{leino2010dafny,
  author       = {K. Rustan M. Leino},
  editor       = {Edmund M. Clarke and
                  Andrei Voronkov},
  title        = {{Dafny: An Automatic Program Verifier for Functional Correctness}},
  booktitle    = {Logic for Programming, Artificial Intelligence, and Reasoning},
  series       = {Lecture Notes in Computer Science},
  volume       = {6355},
  pages        = {348--370},
  publisher    = {Springer},
  year         = {2010},
  url          = {https://doi.org/10.1007/978-3-642-17511-4\_20},
}

@inproceedings{moura2021lean4,
  title={The {Lean 4} theorem prover and programming language},
  author={Moura, Leonardo de and Ullrich, Sebastian},
  booktitle={International Conference on Automated Deduction},
  pages={625--635},
  year={2021},
  organization={Springer}
}

@article{lattuada2023verus,
  title={Verus: Verifying {Rust} programs using linear ghost types},
  author={Lattuada, Andrea and Hance, Travis and Cho, Chanhee and Brun, Matthias and Subasinghe, Isitha and Zhou, Yi and Howell, Jon and Parno, Bryan and Hawblitzel, Chris},
  journal={Proceedings of the ACM on Programming Languages},
  volume={7},
  number={OOPSLA1},
  pages={286--315},
  year={2023},
  publisher={ACM New York, NY, USA}
}

@inproceedings{barbosa2022cvc5,
  author       = {Haniel Barbosa and
                  Clark W. Barrett and
                  Martin Brain and
                  Gereon Kremer and
                  Hanna Lachnitt and
                  Makai Mann and
                  Abdalrhman Mohamed and
                  Mudathir Mohamed and
                  Aina Niemetz and
                  Andres N{\"{o}}tzli and
                  Alex Ozdemir and
                  Mathias Preiner and
                  Andrew Reynolds and
                  Ying Sheng and
                  Cesare Tinelli and
                  Yoni Zohar},
  editor       = {Dana Fisman and
                  Grigore Rosu},
  title        = {cvc5: {A} Versatile and Industrial-Strength {SMT} Solver},
  booktitle    = {Tools and Algorithms for the Construction and Analysis of Systems
                  - 28th International Conference, {TACAS} 2022},
  series       = {Lecture Notes in Computer Science},
  volume       = {13243},
  pages        = {415--442},
  publisher    = {Springer},
  year         = {2022},
}

@inproceedings{barrett2011cvc4,
  title={cvc4},
  author={Barrett, Clark and Conway, Christopher L and Deters, Morgan and Hadarean, Liana and Jovanovi{\'c}, Dejan and King, Tim and Reynolds, Andrew and Tinelli, Cesare},
  booktitle={International Conference on Computer Aided Verification},
  pages={171--177},
  year={2011},
  organization={Springer}
}

@article{padhi2023sygusif2,
  title={The {SyGuS} language standard version 2.1},
  author={Padhi, Saswat and Polgreen, Elizabeth and Raghothaman, Mukund and Reynolds, Andrew and Udupa, Abhishek},
  journal={arXiv preprint arXiv:2312.06001},
  year={2023}
}

@inproceedings{de2008z3,
  title={Z3: An efficient {SMT} solver},
  author={De Moura, Leonardo and Bj{\o}rner, Nikolaj},
  booktitle={International conference on Tools and Algorithms for the Construction and Analysis of Systems},
  pages={337--340},
  year={2008},
  organization={Springer}
}

@book{tla-lang,
  author = {Lamport, Leslie},
  month = {Jun},
  publisher = {Addison-Wesley},
  title = {{Specifying Systems: The TLA+ Language and Tools for Hardware and Software Engineers}},
  year = 2002,
}

@InProceedings{tlc-mc,
author="Yu, Yuan
and Manolios, Panagiotis
and Lamport, Leslie",
editor="Pierre, Laurence
and Kropf, Thomas",
title={{Model Checking TLA+ Specifications}},
booktitle="Correct Hardware Design and Verification Methods",
year="1999",
publisher="Springer Berlin Heidelberg",
address="Berlin, Heidelberg",
pages="54--66",
isbn="978-3-540-48153-9"
}

@software{polysemist-github,
  author       = {Egolf, Derek},
  title        = {{PolySemist TACAS 2025 Artifact}},
  year         = {2025},
  publisher    = {Zenodo},
  doi          = {10.5281/zenodo.14618423},
  url          = {https://doi.org/10.5281/zenodo.14618423}
}

@misc{biere2007aiger,
  title={The {AIGER} and-inverter graph {(AIG)} format version 20071012},
  author={Biere, Armin},
  year={2007}
}

@inproceedings{cavada2014nuxmv,
  title={The {nuXmv} symbolic model checker},
  author={Cavada, Roberto and Cimatti, Alessandro and Dorigatti, Michele and Griggio, Alberto and Mariotti, Alessandro and Micheli, Andrea and Mover, Sergio and Roveri, Marco and Tonetta, Stefano},
  booktitle={International Conference on Computer Aided Verification},
  pages={334--342},
  year={2014},
  organization={Springer}
}

@article{hui2024qwen25coder,
  title={{Qwen2.5-Coder} technical report},
  author={Hui, Binyuan and Yang, Jian and Cui, Zeyu and Yang, Jiaxi and Liu, Dayiheng and Zhang, Lei and Liu, Tianyu and Zhang, Jiajun and Yu, Bowen and Lu, Keming and others},
  journal={arXiv preprint arXiv:2409.12186},
  year={2024}
}

@article{jacobs2016tlsfLanguage,
  title={A high-level {LTL} synthesis format: {TLSF} v1. 1},
  author={Jacobs, Swen and Klein, Felix and Schirmer, Sebastian},
  journal={arXiv preprint arXiv:1604.02284},
  year={2016}
}

@article{jacobs2024SYNTCOMP,
  title={The reactive synthesis competition ({SYNTCOMP}): 2018--2021},
  author={Jacobs, Swen and P{\'e}rez, Guillermo A and Abraham, Remco and Bruyere, Veronique and Cadilhac, Micha{\"e}l and Colange, Maximilien and Delfosse, Charly and van Dijk, Tom and Duret-Lutz, Alexandre and Faymonville, Peter and others},
  journal={International journal on software tools for technology transfer},
  volume={26},
  number={5},
  pages={551--567},
  year={2024},
  publisher={Springer}
}

@article{singh2025gpt5systemcard,
      title={{OpenAI GPT-5 System Card}}, 
  author={Singh, Aaditya and Fry, Adam and Perelman, Adam and Tart, Adam and Ganesh, Adi and El-Kishky, Ahmed and McLaughlin, Aidan and Low, Aiden and Ostrow, AJ and Ananthram, Akhila and others},
  journal={arXiv preprint arXiv:2601.03267},
  year={2025}
}

@misc{cvc5repository,
  title ={{cvc5} Code Repository},
  author={cvc5 Team},
  url={https://github.com/cvc5/cvc5}
}

@online{genaiTLAChallenge2025,
  title        = {GenAI-accelerated TLA\texttt{+} Challenge},
  author       = {{TLA+ Foundation}},
  year         = {2025},
  url          = {https://foundation.tlapl.us/challenge/index.html},
  note         = {Challenge description and results page, accessed 23 January 2026},
  organization = {TLA+ Foundation}
}

@article{alur2019sygusComp2018,
  title={{SyGuS-COMP} 2018: Results and analysis},
  author={Alur, Rajeev and Fisman, Dana and Padhi, Saswat and Singh, Rishabh and Udupa, Abhishek},
  journal={arXiv preprint arXiv:1904.07146},
  year={2019}
}

@misc{brown2020gpt3,
      title={Language Models are Few-Shot Learners}, 
      author={Tom B. Brown and Benjamin Mann and Nick Ryder and Melanie Subbiah and Jared Kaplan and Prafulla Dhariwal and Arvind Neelakantan and Pranav Shyam and Girish Sastry and Amanda Askell and Sandhini Agarwal and Ariel Herbert-Voss and Gretchen Krueger and Tom Henighan and Rewon Child and Aditya Ramesh and Daniel M. Ziegler and Jeffrey Wu and Clemens Winter and Christopher Hesse and Mark Chen and Eric Sigler and Mateusz Litwin and Scott Gray and Benjamin Chess and Jack Clark and Christopher Berner and Sam McCandlish and Alec Radford and Ilya Sutskever and Dario Amodei},
      year={2020},
      eprint={2005.14165},
      archivePrefix={arXiv},
      primaryClass={cs.CL},
      url={https://arxiv.org/abs/2005.14165}, 
}

@inproceedings{acl2s,
 author = {Peter C. Dillinger and
Panagiotis Manolios and
Daron Vroon and
J Strother Moore},
 bibsource = {dblp computer science bibliography, https://dblp.org},
 booktitle = {29th International Conference on Software Engineering {(ICSE} 2007),
Minneapolis, MN, USA, May 20-26, 2007, Companion Volume},
 pages = {59--60},
 publisher = {{IEEE} Computer Society},
 title = {ACL2s: "The {ACL2} Sedan"},
 url = {https://doi.org/10.1109/ICSECOMPANION.2007.14},
 year = {2007}
}

@inproceedings{acl2s-cgen,
 author = {Harsh Raju Chamarthi and
Peter C. Dillinger and
Matt Kaufmann and
Panagiotis Manolios},
 bibsource = {dblp computer science bibliography, https://dblp.org},
 booktitle = {Proceedings 10th International Workshop on the {ACL2} Theorem Prover
and its Applications, {ACL2} 2011, Austin, Texas, USA, November 3-4,
2011},
 editor = {David S. Hardin and
Julien Schmaltz},
 pages = {4--19},
 series = {{EPTCS}},
 title = {Integrating Testing and Interactive Theorem Proving},
 url = {https://doi.org/10.4204/EPTCS.70.1},
 volume = {70},
 year = {2011}
}

@article{clarkson2010hyperproperties,
  title={Hyperproperties},
  author={Clarkson, Michael R and Schneider, Fred B},
  journal={Journal of Computer Security},
  volume={18},
  number={6},
  pages={1157--1210},
  year={2010},
  publisher={SAGE Publications Sage UK: London, England}
}

@inproceedings{barrett2010smtLib,
  title={The {SMT-LIB} standard: Version 2.0},
  author={Barrett, Clark and Stump, Aaron and Tinelli, Cesare and others},
  booktitle={Proceedings of the 8th international workshop on satisfiability modulo theories (Edinburgh, UK)},
  volume={13},
  pages={14},
  year={2010}
}

@inproceedings{contata,
 author = {Anders Miltner and
Ziteng Wang and
Swarat Chaudhuri and
Isil Dillig},
 bibsource = {dblp computer science bibliography, https://dblp.org},
 booktitle = {Computer Aided Verification - 36th International Conference, {CAV}
2024, Montreal, QC, Canada, July 24-27, 2024, Proceedings, Part {III}},
 editor = {Arie Gurfinkel and
Vijay Ganesh},
 pages = {41--63},
 publisher = {Springer},
 series = {Lecture Notes in Computer Science},
 title = {Relational Synthesis of Recursive Programs via Constraint Annotated
Tree Automata},
 url = {https://doi.org/10.1007/978-3-031-65633-0\_3},
 volume = {14683},
 year = {2024}
}

@article{burst,
 author = {Anders Miltner and
Adrian Trejo Nu{\~{n}}ez and
Ana Brendel and
Swarat Chaudhuri and
Isil Dillig},
 bibsource = {dblp computer science bibliography, https://dblp.org},
 journal = {Proc. {ACM} Program. Lang.},
 number = {{POPL}},
 pages = {1--29},
 title = {Bottom-up synthesis of recursive functional programs using angelic
execution},
 url = {https://doi.org/10.1145/3498682},
 volume = {6},
 year = {2022}
}

@inproceedings{synquid,
 author = {Nadia Polikarpova and
Ivan Kuraj and
Armando Solar{-}Lezama},
 bibsource = {dblp computer science bibliography, https://dblp.org},
 booktitle = {Proceedings of the 37th {ACM} {SIGPLAN} Conference on Programming
Language Design and Implementation, {PLDI} 2016, Santa Barbara, CA,
USA, June 13-17, 2016},
 editor = {Chandra Krintz and
Emery D. Berger},
 pages = {522--538},
 publisher = {{ACM}},
 title = {Program synthesis from polymorphic refinement types},
 url = {https://doi.org/10.1145/2908080.2908093},
 year = {2016}
}

@inproceedings{leon,
 author = {Etienne Kneuss and
Ivan Kuraj and
Viktor Kuncak and
Philippe Suter},
 bibsource = {dblp computer science bibliography, https://dblp.org},
 booktitle = {Proceedings of the 2013 {ACM} {SIGPLAN} International Conference on
Object Oriented Programming Systems Languages {\&} Applications,
{OOPSLA} 2013, part of {SPLASH} 2013, Indianapolis, IN, USA, October
26-31, 2013},
 editor = {Antony L. Hosking and
Patrick Th. Eugster and
Cristina V. Lopes},
 pages = {407--426},
 publisher = {{ACM}},
 title = {Synthesis modulo recursive functions},
 url = {https://doi.org/10.1145/2509136.2509555},
 year = {2013}
}

@article{para,
 author = {Qiantan Hong and
Alex Aiken},
 bibsource = {dblp computer science bibliography, https://dblp.org},
 journal = {Proc. {ACM} Program. Lang.},
 number = {{PLDI}},
 pages = {102--125},
 title = {Recursive Program Synthesis using Paramorphisms},
 url = {https://doi.org/10.1145/3656381},
 volume = {8},
 year = {2024}
}

@inproceedings{escher,
 author = {Aws Albarghouthi and
Sumit Gulwani and
Zachary Kincaid},
 bibsource = {dblp computer science bibliography, https://dblp.org},
 booktitle = {Computer Aided Verification - 25th International Conference, {CAV}
2013, Saint Petersburg, Russia, July 13-19, 2013. Proceedings},
 editor = {Natasha Sharygina and
Helmut Veith},
 pages = {934--950},
 publisher = {Springer},
 series = {Lecture Notes in Computer Science},
 title = {Recursive Program Synthesis},
 url = {https://doi.org/10.1007/978-3-642-39799-8\_67},
 volume = {8044},
 year = {2013}
}

@inproceedings{myth,
 author = {Peter{-}Michael Osera and
Steve Zdancewic},
 bibsource = {dblp computer science bibliography, https://dblp.org},
 booktitle = {Proceedings of the 36th {ACM} {SIGPLAN} Conference on Programming
Language Design and Implementation, Portland, OR, USA, June 15-17,
2015},
 editor = {David Grove and
Stephen M. Blackburn},
 pages = {619--630},
 publisher = {{ACM}},
 title = {Type-and-example-directed program synthesis},
 url = {https://doi.org/10.1145/2737924.2738007},
 year = {2015}
}

@article{smyth,
 author = {Justin Lubin and
Nick Collins and
Cyrus Omar and
Ravi Chugh},
 bibsource = {dblp computer science bibliography, https://dblp.org},
 journal = {Proc. {ACM} Program. Lang.},
 number = {{ICFP}},
 pages = {109:1--109:29},
 title = {Program sketching with live bidirectional evaluation},
 url = {https://doi.org/10.1145/3408991},
 volume = {4},
 year = {2020}
}

@inproceedings{lambda-sq,
 author = {John K. Feser and
Swarat Chaudhuri and
Isil Dillig},
 bibsource = {dblp computer science bibliography, https://dblp.org},
 booktitle = {Proceedings of the 36th {ACM} {SIGPLAN} Conference on Programming
Language Design and Implementation, Portland, OR, USA, June 15-17,
2015},
 editor = {David Grove and
Stephen M. Blackburn},
 pages = {229--239},
 publisher = {{ACM}},
 title = {Synthesizing data structure transformations from input-output examples},
 url = {https://doi.org/10.1145/2737924.2737977},
 year = {2015}
}

\newpage
\appendix

\section{Prompt Templates}
\label{sec_appendix_prompts}
\subsection{Prompts for LTL Synthesis with AIGER Output}
\label{app:ltl_synth_prompt_AIGER}

\begin{lstlisting}[style=systempromptSimple]
You are an expert LTL synthesis engine. Your task is to synthesize a circuit in ASCII AIGER format
from a given TLSF specification. The circuit must satisfy all the LTL guarantees specified in the TLSF.

## ASCII AIGER Format Overview

AIGER (And-Inverter Graph) is a circuit representation using only AND gates and inverters.

**Header: `aag M I L O A`**
- M = maximum variable index (total variables used)
- I = number of inputs
- L = number of latches (state elements)
- O = number of outputs
- A = number of AND gates

**Structure:**
```
aag M I L O A
<inputs>      -- I lines
<latches>     -- L lines (if L > 0)
<outputs>     -- O lines
<AND gates>   -- A lines
<symbols>     -- Optional
c
<comments>    -- Optional
```

## Literal Encoding

Literals: **even = positive**, **odd = negated**
- Literal 2 = variable 1 (positive), Literal 3 = variable 1 (negated)
- Literal 4 = variable 2 (positive), Literal 5 = variable 2 (negated)
- **Variable index = literal / 2**
- FALSE = 0, TRUE = 1

## Variable Ordering

Variables numbered in order: inputs (1..I) -> latches (I+1..I+L) -> AND gates (I+L+1..M)

## AND Gate Definition

Format: `lhs rhs0 rhs1`
- `lhs` must be **EVEN** (defines new variable)
- Gate computes: `lhs_var = rhs0 AND rhs1`

## Latches (Sequential Logic)

Format: `<current> <next>` - latch initialized to 0, next state from literal.

Example: `4 8` -> current state is var2 (literal 4), next from gate 8's output.

## Input/Output Symbol Constraints (CRITICAL)

Your AIGER circuit MUST have input/output symbols that **EXACTLY match** the TLSF specification:

1. **Number of signals**: Input/output count in AIGER must match TLSF INPUTS/OUTPUTS
2. **Signal names**: Symbol names must be **IDENTICAL** to TLSF variable names
3. **Example**: If TLSF has `INPUTS { a; b; }` and `OUTPUTS { outp; }`:
   - AIGER must have exactly 2 inputs with symbols `i0 a` and `i1 b`
   - AIGER must have exactly 1 output with symbol `o0 outp`

**Do NOT invent generic names** - use the EXACT names from TLSF.

## Example: TLSF to AIGER

**Input TLSF** (`trivial_and.tlsf`):
```
INFO {
    TITLE:       "Trivial AND gate"
    DESCRIPTION: "Output is AND of two inputs"
    SEMANTICS:   Mealy
    TARGET:      Mealy
}
MAIN {
    INPUTS {
        a;
        b;
    }
    OUTPUTS {
        outp;
    }
    GUARANTEES {
        G(outp <-> (a && b));
    }
}
```

**Output AIGER**:
```
REALIZABLE
aag 3 2 0 1 1
2
4
6
6 2 4
i0 a
i1 b
o0 outp
```

**Key points**:
- Header: `aag 3 2 0 1 1` -> M=3 (max var), I=2 inputs, L=0 latches, O=1 output, A=1 AND gate
- Inputs: `2` (literal for a), `4` (literal for b)
- Output: `6` (literal for outp, referencing AND gate result)
- AND gate: `6 2 4` -> var3 = var1 AND var2
- Symbols: `i0 a`, `i1 b`, `o0 outp` - exactly match TLSF INPUTS/OUTPUTS

## Common Mistakes to Avoid

1. **Wrong header counts** - M, I, L, O, A must match actual content
2. **Odd literals for AND outputs** - LHS must be EVEN
3. **Wrong symbol names** - Must exactly match TLSF INPUTS/OUTPUTS
4. **Missing REALIZABLE header** - First line must be "REALIZABLE"

## Output Format

Your output MUST:
1. Start with "REALIZABLE"
2. Follow with valid AAG header starting with "aag"
3. Include symbols matching TLSF signal names exactly
4. Use proper literal encoding (even=positive, odd=negated)
5. NO explanations, NO commentary - ONLY the circuit
\end{lstlisting}

\begin{lstlisting}[style=userpromptSimple]
Synthesize the solution in ASCII AIGER format for the following TLSF specification:

```tlsf
{{ tlsf_content }}
```

Output must start with "REALIZABLE" followed by a valid AIGER circuit that satisfies all LTL guarantees.
\end{lstlisting}

\subsection{Prompts for LTL Synthesis with SMV Output}
\label{app:ltl_synth_prompt_SMV}
\begin{lstlisting}[style=systempromptSimple]
You are an expert LTL (Linear Temporal Logic) synthesis tool which solves synthesis problems like those in The Reactive Synthesis Competition (SYNTCOMP).

Your Goal: Generate a deterministic finite-state machine in nuXmv SMV format that satisfies the given LTL specification.

Input: A TLSF specification defining Inputs, Outputs, Assumptions, Definitions, Invariants, LTL Guarantees, and so on.

Output Format: nuXmv module.

Strict Structure Constraints:

1. Module Name: Must be MODULE main

2. IVAR Section:
   - Declare all input variables (from TLSF INPUTS) with type ": boolean;"
   - DO NOT use integer, enum, or array types - only boolean variables
   - Example:
     IVAR
       i1 : boolean;
       i2 : boolean;
   
3. VAR Section:
   - Declare any internal state variables (if needed) with type ": boolean;"
   - DO NOT use integer, enum, or array types - only boolean variables
   - Example:
     VAR
       state1 : boolean;
       state2 : boolean;
   
4. ASSIGN Section:
   - For each internal state variable x, add a single "init(x) := ... ;" initialization statement in the ASSIGN section: fill in the "..." part with a boolean expression. 
   - For each internal state variable x, add a single "next(x) := ... ;" transition function statement in the ASSIGN section: fill in the "..." part with a boolean expression.
   - Example:
     ASSIGN
       init(state1) := FALSE;
       init(state2) := TRUE;
       next(state1) := i1 & state1; 
       next(state2) := i2 | state1;

5. DEFINE Section:
   - For each output variable y (from TLSF OUTPUTS), add a single "y := ... ;" statement defining that output variable: fill in the "..." part with a boolean expression.
   - Example:
     DEFINE
       out := i2 | state2;
 
IMPORTANT Requirements:
  
1. IMPORTANT: do NOT include any INIT or TRANS sections. ONLY include a single IVAR section, a single VAR section, a single ASSIGN section and a single DEFINE section.

2. Boolean Operators:
   NOT: !         (important: use ! not ~)
   AND: &         (important: use & not &&)
   OR: |          (important: use | not ||)
   Implication: ->
   Iff: <->

   IMPORTANT: nuXmv uses single-character operators, NOT double-character:
   - Use ! for NOT (not ~)
   - Use & for AND (not &&)
   - Use | for OR (not ||)

   Valid: !a & b | c
   Invalid: ~a && b || c (~, &&, || are invalid)

   NOTE: Do NOT use IF-THEN-ELSE-FI syntax. Use boolean expressions with !, &, | for conditionals.

3. do NOT add the LTLSPEC statements, as the system will add them from the TLSF automatically

4. input variables must ONLY be declared in the IVAR section. Do not declare or define them elsewhere.

5. state variables (if needed) must ONLY be declared in the VAR section.

6. output variables (from OUTPUTS section of TLSF specification) must ONLY be defined in the DEFINE section. do NOT declare them in IVAR or VAR.

7. do NOT assign values to input variables. they are controlled by the environment, not the system. In particular, NEVER use init(input_var) or next(input_var) in the ASSIGN section.

8. the ASSIGN section is exclusively for initializing and updating state variables declared in VAR. Every variable declared in VAR MUST have exactly one init(...) statement and exactly one next(...) statement in the ASSIGN section.

9. if a variable is declared in IVAR or VAR, it must NOT appear on the left-hand side of any DEFINE statement. Only output variables (which are not declared elsewhere) may be defined in the DEFINE section.

10. output ONLY the raw SMV code inside a markdown code block

Example Structure:
```smv
MODULE main
IVAR
  -- inputs
  input1 : boolean;
  input2 : boolean;
VAR
  -- state variables (if needed)
  state1 : boolean;
  state2 : boolean;
ASSIGN
  -- your init and next assignments here
  init(state1) := ... ;
  init(state2) := ... ;
  next(state1) := ... ;
  next(state2) := ... ;
DEFINE
  -- your output assignments here
  output1 := ... ;
  output2 := ... ;
```


\end{lstlisting}

\begin{lstlisting}[style=userpromptSimple]
Synthesize the solution in nuXmv SMV format for the following TLSF specification:

```tlsf
{{ tlsf_content }}
```
\end{lstlisting}

\subsection{Prompts for SyGuS Synthesis}
\label{app:sygus_synth_prompt}
\begin{lstlisting}[style=systempromptSimple]
You are an expert in program synthesis and SMT-LIB/SyGuS format.

Your task is to synthesize a function in SMT-LIB format that satisfies the given SyGuS specification.

IMPORTANT RULES:
1. Output ONLY the (define-fun ...) expression(s) that solve the synthesis problem
2. The function must satisfy ALL constraints in the specification
3. If a grammar is specified in the synth-fun, your solution MUST use only the allowed operations from that grammar. Do not use operators outside the grammar even if they seem necessary.
4. **CRITICAL: The define-fun signature MUST match the synth-fun signature EXACTLY**
   - If synth-fun has NO parameters, define-fun must have NO parameters
   - If synth-fun has parameters, define-fun must have the SAME parameters with SAME types
   - DO NOT add parameters from grammar nonterminals (like Start, StartBool) to your define-fun
5. Do NOT include (assume ...) or (declare-var ...) commands in your output - these are not part of the solution
6. Do NOT define auxiliary/helper functions. Logic must be self-contained within the requested function.
7. Do NOT include any explanation - output only the SMT-LIB code
8. CRITICAL: Use ONLY standard SMT-LIB primitives. For integer arithmetic (LIA): +, -, *, div, mod, comparisons (=, <, <=, >, >=), ite, and, or, not. For negative numbers, use unary negation (- 5) instead of -5 if required by strict SMT-LIB compliance.
   WARNING: strictly avoid `max`, `min`, `abs`, `if`, `return`. Use `ite` for conditionals (not `if`). Implement `max`, `min`, `abs` using ite and comparisons.
9. For bitvectors (BV): bvand, bvor, bvxor, bvnot, bvadd, bvsub, bvmul, bvudiv, bvurem, bvshl, bvlshr, bvult, bvule, bvugt, bvuge, concat, extract
   Ensure literals are typed correctly, e.g., use `#x0001` or `(_ bv<value> <width>)`, where <value> is a decimal value, <width> is the bitwidth
10. For strings: str.++ (concatenation), str.len, str.at, str.substr, str.prefixof, str.suffixof, str.contains, str.indexof, str.replace, str.to.int, int.to.str
11. For arrays: store, select
12. For Booleans: Use true and false (lowercase). Do NOT use 1 or 0 for boolean values.

EXAMPLE:

Given this SyGuS specification:
```sygus
(set-logic LIA)

(synth-fun mi ((x Int) (y Int)) Int
    ((S Int) (I Int))
    ((S Int ((ite (<= I I) I I)))
     (I Int (x y))))

(declare-var x Int)
(declare-var y Int)
(constraint (=> (>= x y) (= (mi x y) y)))
(constraint (=> (>= y x) (= (mi x y) x)))

(check-synth)
```

The correct output is:
```smt2
(define-fun mi ((x Int) (y Int)) Int (ite (<= x y) x y))
```

Note that:
- The function signature `(define-fun mi ((x Int) (y Int)) Int ...)` matches `(synth-fun mi ((x Int) (y Int)) Int ...)`
- The grammar nonterminals (S, I) are NOT in the function signature
- The solution satisfies the constraint: `(=> (>= x y) (= (mi x y) y))` and `(=> (>= y x) (= (mi x y) x))`
- The solution body is `(ite (<= x y) x y)`, conforming to the grammar

OUTPUT FORMAT:
```smt2
(define-fun <name> (<args>) <return-type> <body>)
(define-fun <name2> (<args>) <return-type> <body>)
...
```

If there are multiple synth-fun declarations, output all of them.

Remember:
- DO NOT output (assume ...) commands
- DO NOT output inv-constraint (if present)
- DO NOT output grammar definitions
- ONLY output the (define-fun ...) expressions
- Function signature MUST match synth-fun signature EXACTLY
\end{lstlisting}

\begin{lstlisting}[style=userpromptSimple]
Synthesize a SMT-LIB function for the following SyGuS specification:

```sygus
{{ sygus_content }}
```

IMPORTANT:
- Output ONLY the (define-fun ...) expression(s) that satisfy all constraints.
- Strictly adhere to the grammar defined in the specification (if present).
\end{lstlisting}

\subsection{Prompts for \TLA Distributed Protocol Synthesis}
\label{app:tla_synth_prompt}

\begin{lstlisting}[basicstyle=\small\ttfamily]
You are an expert at writing TLA+ protocols. You must  
fill the holes {{ list_of_holes }} in {{ list_of_actions }}
in the following sketch:

{{ sketch }}

Substituting the mapping into the sketch must produce a 
TLA+ protocol that satisfies the properties: 

{{ properties }}

For each hole, you must use the associated grammar. Below
the grammars are shown. You must use parentheses only as 
specified by the grammars. Omitting or adding them  
inappropriately is not acceptable.

{{ grammars }}

Do not output prose. Do not explain. Do not output code 
fences. Do not output raw TLA+ code. Only output a json
structure with the appropriate keys. The json syntax must 
be correct; use string escapes where necessary.

Your final output must be a json mapping of the form:

{{ json_map_structure }}
\end{lstlisting}

\subsection{Prompts for Recursive Program Synthesis}
\label{app:acl2s_synth_prompt}

\begin{lstlisting}[basicstyle=\small\ttfamily]

You are an expert at writing code in the lisp-like acl2s 
programming language.

Write function(s) with the following template(s) (fill in 
<BODY>):

{{ sig_block }}

You may use the following primitive functions:

{{ primitives_block }}

And you may apply these functions to these terminals 
(constants and function arguments):

{{ terminals_block }}

Each function may call itself and it may use if statements.
If multiple functions, they may call each other.

Your function(s) must satisfy the following properties:

{{ properties_block }}

Your function(s) must satisfy the following input-output 
example(s):

{{ io_block }}

The following declarations define (mutually) inductive 
datatypes and expose their constructors/destructors:

{{ datatype_block }}

It may be useful to know how some data types and functions
are defined:

{{ definitions_block }}

Sketch guidance: For each function listed below, you may 
choose any one of its sketches to complete.

{{ sketches_block }}

The following grammar defines the valid expressions for 
filling holes:

{{ grammar_block }}

Implement the function(s) now. Do not include any prose, 
explanations, or code fences. Just provide the definitions 
of the new function(s): {{ function_list }}.

\end{lstlisting}

\section{Determinism and Sanity Checks for LLM Generation of SMV Output}
\label{app:ltl_synth_sanity}
\subsection{Determinism Checking by Self-Composition}

We verify the determinism of the LLM-generated \nuxmv modules by leveraging the concept of self-composition, a standard technique for verifying hyperproperties~\cite{clarkson2010hyperproperties}. Since determinism requires that a system produces identical behaviors for identical input sequences, we use self-composition to compare two instances of the module under the same inputs.

The procedure, which we explain through the example in Figure~\ref{fig:full_determinism_check}, is as follows:
\begin{enumerate}
    \item \textbf{Module Transformation:} We transform the candidate module (e.g., Figure~\ref{fig:ltl_self_compose_example_module}) into a sub-module (e.g. in Figure~\ref{fig:ltl_self_compose_example_check}). This involves renaming \texttt{MODULE main} to a unique name (e.g., \texttt{testmod}) and lifting all declarations from the \texttt{IVAR} section into the module's input.
    \item \textbf{Self Composition:} We create a new \texttt{main} module that declares the global input variables and instantiates two distinct copies of the transformed module, denoted as $t_1$ and $t_2$ (see Figure~\ref{fig:ltl_self_compose_example_check}).
    \item \textbf{Input Synchronization:} We bind the same global input variables to both instances, ensuring they receive identical input sequences.
    \item \textbf{Determinism Check:} We verify a CTL specification asserting that for all reachable states (\texttt{AG}), the internal state variables of $t_1$ and $t_2$ remain equal. If the model checker confirms this property, the module is deterministic; otherwise, a counterexample demonstrates a sequence of inputs leading to divergent states.
\end{enumerate}

This check also catches cases where initial values or next-state updates are undefined, which lead to non-determinism.

\begin{figure}[ht]
    \centering
    \begin{subfigure}[t]{0.42\textwidth}
        \begin{lstlisting}[basicstyle=\scriptsize\ttfamily, frame=single, xleftmargin=2pt]
MODULE main
IVAR
  i : boolean;
  j : boolean;
VAR
  x : boolean;
  y : boolean;
ASSIGN
  init(x) := FALSE; 
  next(x) := i & y; 
  init(y) := FALSE; 
  next(y) := !y | j;
DEFINE
  out := x & y & i & j;
        \end{lstlisting}
        \caption{Candidate \nuxmv module generated by the LLM.}
        \label{fig:ltl_self_compose_example_module}
    \end{subfigure}
    \hfill 
    \begin{subfigure}[t]{0.54\textwidth}
        \begin{lstlisting}[basicstyle=\scriptsize\ttfamily, frame=single, xleftmargin=2pt]
MODULE testmod(i, j)
VAR
  x : boolean;
  y : boolean;
ASSIGN
  init(x) := FALSE; 
  next(x) := i & y; 
  init(y) := FALSE; 
  next(y) := !y | j;
DEFINE
  out := x & y & i & j;

MODULE main
IVAR
  i : boolean;
  j : boolean;
VAR
  t1 : testmod(i, j);
  t2 : testmod(i, j);
DEFINE
  -- Check internal states match
  state_eq := (t1.x = t2.x) & 
              (t1.y = t2.y);

CTLSPEC AG state_eq
        \end{lstlisting}
        \caption{Self-composed new module and the CTL specification for checking determinism.}
        \label{fig:ltl_self_compose_example_check}
    \end{subfigure}
    
    \caption{Checking determinism through self-composition.}
    \label{fig:full_determinism_check}
\end{figure}

\subsection{Other Sanity Checks}

To ensure the validity of the LLM-generated \nuxmv modules for LTL synthesis, we implemented the following additional sanity checks during the post-synthesis verification phase:
\begin{itemize}
    \item \textbf{Structure Check:} 
    Adhering to the prompt instructions (Appendix~\ref{app:ltl_synth_prompt_SMV}), the generated module must contain specific sections: \texttt{IVAR} and \texttt{DEFINE} to specify inputs and outputs, while \texttt{VAR} and \texttt{ASSIGN} are included only if state variables are generated. We parse the generated code to verify the presence of these sections; missing sections flag the module as incomplete or malformed. 
    Additionally, this check confirms the absence of \texttt{TRANS} and \texttt{INIT} sections, which are explicitly prohibited by the prompt instructions. 
    \item \textbf{Boolean Variable Check:}
    We ensure that all input and state variables are declared as \texttt{boolean}, as the prompt instructs.
    \item \textbf{Input/Output Mapping Check:} 
    The \nuxmv module must use the exact input and output identifiers defined in the TLSF specification. We extract and compare the variable names from both the specification and the generated module. Any discrepancy is flagged as a failure to adhere to the interface.
\end{itemize}

In our experiments, no structural issues were detected across any \qwenSMV or \gptSMV correct solutions. 
We also found that all variables are correctly declared as \texttt{boolean} for all the correct solutions generated by \qwenSMV and \gptSMV.
Regarding input/output mapping, the 229 correct \gptSMV solutions showed perfect alignment with the specifications. However, among the 50 correct \qwenSMV solutions, we identified 2 cases where the variable names did not match the TLSF specification. We manually inspected these cases and confirmed that, despite the mismatch, the logic was semantically correct. Therefore, we include them in the count of successful solutions.

\end{document}